\documentstyle[10pt,aas2pp4]{article}

\def\micron{\,$\mu$m}
\def\etal{{et~al.}}

\def\sec{\prime\prime}
\def\min{\prime}

\def\Hone{H{\sc i}}

\def\farcs{\hbox{$\> .\!\!^{\prime\prime}$}}

\begin{document}

\title{Extinction Curves, Distances, and Clumpiness of Diffuse Interstellar
Dust Clouds\footnote{Observations carried out at the Cerro Tololo
Interamerican Observatory, National Optical Astronomy Observatories, which is
operated by the Association of Universities for Research in Astronomy, Inc.\
under cooperative agreement with the National Science Foundation.}}

\author{Arpad Szomoru}
\and
\author{Puragra Guhathakurta\altaffilmark{2}}
\affil{UCO/Lick Observatory, University of California, Santa Cruz,
California~95064, USA}
\authoremail{arpad@ucolick.org}
\authoremail{raja@ucolick.org}
\altaffiltext{2}{Alfred P. Sloan Research Fellow}

\begin{abstract} 

We present CCD photometry in $UBVRI$ of several thousand Galactic field stars
in four~large ($>1$~degree$^2$) regions centered on diffuse interstellar dust
clouds, commonly referred to as ``cirrus'' clouds (with optical depth $A_V$
less than unity).  Our goal in studying these stars is to investigate the
properties of the cirrus clouds.  A comparison of the observed stellar
surface density between on-cloud and off-cloud regions as a function of
apparent magnitude in each of the five~bands effectively yields a measure of
the extinction through each cloud.  For two of the cirrus clouds, this method
is used to derive $UBVRI$ star counts-based extinction curves, and $U$-band
counts are used to place constraints on the cloud distance.  The color
distribution of stars and their location in ($U-B$,~$B-V$) and ($B-V$,~$V-I$)
color-color space are analyzed in order to determine the amount of selective
extinction (reddening) caused by the cirrus.  The color excesses,
$A_\lambda-A_V$, derived from stellar color histogram offsets for the
four~clouds, are better fit by a reddening law that rises steeply towards
short wavelengths [$R_V\equiv{A}_V/E(B-V)\lesssim2$] than by the standard law
($R_V=3.1$).  This may be indicative of a higher-than-average abundance of
small dust grains relative to larger grains in diffuse cirrus clouds.  The
shape of the counts-based effective
extinction curve and a comparison of different estimates of the dust optical
depth (extinction optical depth derived from background star counts/colors;
emission optical depth derived from far infrared measurements), are used to
measure the degree of clumpiness in clouds.  The set of techniques explored
in this paper can be readily adapted to the Sloan Digital Sky Survey data set
in order to carry out a systematic, large-scale study of cirrus clouds. 

\end{abstract}


\section{Introduction}
\label{intro}

Among the most remarkable of the many discoveries made by the {\it Infrared
Astronomical Satellite\/} ({\it IRAS\/}, \markcite{neugebauer84}Neugebauer
\etal\ 1984) was the detection of a bright component of the far infrared
background emission.  This far infrared component is marked by discrete
condensations of filamentary clouds with characteristic sizes ranging between
a few and $\sim30^\circ$.  Their resemblance to clouds in our own atmosphere
led \markcite{low84}Low \etal\ (1984) to dub them the infrared ``cirrus''.
The cirrus clouds are most readily observed at high Galactic latitudes where
there is relatively little contribution to the net background by the Galactic
plane, and are most prominent at 100\micron, although they are visible in all
four~{\it IRAS\/} bands.  Low \etal\ noted the positional correlation of some
of the cirrus emission with {\Hone} clouds mapped by
\markcite{heiles75}Heiles (1975) and interpreted this as evidence that the
infrared cirrus was in fact associated with the diffuse interstellar medium
(ISM).

Despite the fact that optical cirrus was first discovered over four~decades
ago \markcite{devau55}\markcite{devau60}(de~Vaucouleurs 1955, 1960), definite
association with high latitude dust came much later
\markcite{devau72}\markcite{sandage76}(de~Vaucouleurs \& Freeman 1972;
Sandage 1976).  The first systematic comparison between the {\it IRAS\/}
100$\,$\micron\ and optical morphologies of dust clouds was conducted by
\markcite{devries85}de~Vries \& Le~Poole (1985).  This and a few other
studies \markcite{paley90}\markcite{stark93}(Paley 1990; Stark 1993), all
based on photographic data, reveal a good correspondence between optical and
infrared cirrus.  Quantitative optical surface brightness measurements of
cirrus clouds have been carried out by \markcite{guha89}Guhathakurta \&
Tyson (1989), \markcite{paley91}Paley \etal\ (1991), and
\markcite{gordon98}Gordon, Witt, \& Friedmann (1998), while spectroscopy of
diffuse cirrus clouds has been carried out only recently
\markcite{szomoru98}(Szomoru \& Guhathakurta 1998).  Such optical
measurements provide important clues towards understanding grain optical
properties, such as the albedo and the phase function asymmetry, and have
demonstrated the existence of extended red emission, caused by
photoluminscence in very small grains, in the diffuse ISM
\markcite{szomoru98}\markcite{gordon98}(Szomoru \& Guhathakurta 1998; Gordon
\etal\ 1998). 

Diffuse cirrus clouds typically extend over large areas on the sky.  Such
clouds can be probed through the study of their effect on the light of
background sources (Galactic stars, distant quasars).  A statistical method
which is well suited for cirrus clouds employs star counts to determine the
extinction.  This method, which was first described and applied several
decades ago \markcite{wolf23}\markcite{bok37}(Wolf 1923; Bok 1937), has been
used to determine the extinction in high latitude molecular clouds
\markcite{magnani86}\markcite{stark95}(cf.~Magnani \& de~Vries 1986; Stark
1995).  By comparing cumulative star counts in the direction of an
interstellar cloud with star counts in extinction-free regions, or
alternatively with model-based star counts (so-called Wolf
diagrams---\markcite{wolf23}Wolf 1923), one may derive both the amount of
extinction through and the distance to the cloud.  An alternate method relies
on optical and ultraviolet spectroscopy of individual background hot stars of
known spectral type \markcite{stark95}(Stark 1995) or QSOs with mostly
featureless, power-law spectra \markcite{bowen91}\markcite{bowen95}(cf.~Bowen
1991; Bowen, Blades, \& Pettini 1995).  This latter method is useful for the
study of dust extinction curves, cloud chemical abundances, and excitation
conditions in the gaseous ISM.  It has been applied mostly to dense molecular
clouds cores with $A_V>\!\!>1$ and to the warm ionized ISM of the Galaxy,
rather than to cirrus clouds. 

This paper presents photometry in $UBVRI$ of several thousand stars in
four~large area ($>1$~degree$^2$) fields centered on cirrus clouds.  The
surface photometry of these clouds will be the subject of a future paper.
Our study differs from previous studies in several respects:

\begin{itemize}
\item{CCD photometry of faint stars over a wide field is less subject to
systematic error than photographic measurements thanks in large part to
accurate flat fielding and sky subtraction.}

\item{The use of $UBVRI$ bands opens up the possibility of studying the
effect of selective extinction by interstellar dust on stellar color
distributions and color-color diagrams, in addition to its effect on star
counts.  This combination of techniques can be used to investigate the
occurence of dense cores and to quantify the amount of structure within the
clouds (Sec.~\ref{clumpy}).}

\item{Measurements of the effect of extinction on star counts/color
distribution are made relative to a well-matched off-cirrus sample of stars,
and this yields a reliable and precise determination of the extinction curve
($A_\lambda$).}

\item{Data at short wavelengths (3600\AA\ $U$ band) enables us to study
clouds of relatively low optical depth; all four~clouds studied here have
$A_V$ values significantly less than unity.  Most previous studies employing
background star counts have been of molecular clouds with higher optical
extinction than the cirrus clouds in our sample, with the study by
\markcite{stark95}Stark (1995) being a notable exception; in fact, two of our
four~clouds are also in his sample.  In contrast to our $UBVRI$ CCD data,
however, the optical part of his study is based solely on $U$-band
photographic plates.} 
\end{itemize}

Direct measurements of the optical extinction have a number of important
applications.  Firstly, the ratio of $A_V$ (visual extinction optical depth)
to $\tau_{100\mu\rm{m}}$ (the far infrared emission optical depth) provides a
test, albeit a crude one, of interstellar dust grain models.  Such tests are
now possible thanks to accurate $A_V$ measurements, along with recent
improvements in techniques for the processing of far infrared data and hence
in the measurement of $\tau_{100\mu\rm{m}}$
\markcite{schlegel98}\markcite{verter98a}\markcite{verter98b}(Schlegel,
Finkbeiner, \& Davis 1998, hereafter SFD; Verter \& Rickard 1998; Verter
\etal\ 1998).  Secondly, a direct measurement of $A_V$ (as opposed to an
approximate scaling based on the far infrared surface brightness) is
essential in order to model the degree of penetration and internal reddening
of ambient starlight in cirrus clouds; this is a key piece of information
used in the interpretation of ultraviolet/optical/near-infrared spectra and
$UBVRI$ surface brightness measurements of optical cirrus. 

The observations and reduction of optical data are described in
Sec.~\ref{optdata} and far infrared measurements are described in
Sec.~\ref{irdata}.  Sec.~\ref{counts} contains the background star count
analysis along with the resulting determination of the extinction law and
cloud distance; Sec.~\ref{colors} contains a description of stellar
color-color diagrams and their use in the measurement of selective extinction
(reddening).  In Sec.~\ref{discuss}, the optical extinction is compared to
the 100\micron\ optical depth, and optical depth estimates from the various
methods are discussed in the context of cloud structure (i.e.,~clumpiness)
and dust properties.  Sec.~\ref{summary} contains a summary of the main
results.

\section{Optical Data}
\label{optdata}

\subsection{Targets and Observations}
\label{targetsobs}

Over the last decade, several compact, relatively isolated, high
latitude cirrus clouds have been observed at optical wavelengths for the
purpose of studying the reprocessing of starlight by interstellar dust
grains \markcite{guha94}(cf.~Guhathakurta \& Cutri 1994).  These cirrus
targets have been selected from {\it IRAS\/} 100\micron\ maps or drawn
from optical cirrus catalogs such as the list of Lynds Bright Nebulae
\markcite{lynds65}(Lynds 1965) or the compilation of
\markcite{paley90}Paley (1990) which are, in turn, derived from Palomar
Observatory Sky Survey plates.  This paper is based on observations of
four~such cirrus clouds, RCrA (R~Corona Australis), PV\,1 (Project
Verification~\#1, an early-phase target of the Infrared Space Observatory),
Paley\,1, and Paley\,3 (from the \markcite{paley90}Paley 1990 catalog).

The observations were carried out with the Cerro Tololo Interamerican
Observatory\footnote{Cerro Tololo Interamerican Observatory, National Optical
Astronomy Observatories, is operated by the Association of Universities for
Research in Astronomy, Inc.\ under cooperative agreement with the National
Science Foundation.} Curtis-Schmidt telescope, equipped with a Tektronix
$2048\times2048$ CCD camera, during three~nights in 1995~October.  This
telescope has a 0.9-m primary mirror and a 0.6-m corrector.  The pixel size
of 21\micron, corresponding to a scale of $2\farcs03$~pixel$^{-1}$, yields a
field of view of $1.15^\circ\times1.15^\circ$.  The finite filter size causes
slight vignetting ($\sim20\%$ at the extreme corners of each CCD frame) but
this is well corrected by sky flat fields. 

The Schmidt CCD observations consist of a series of disregistered 400~s
exposures in the standard $UBVRI$ filter set at Cerro Tololo Interamerican
Observatory.  The effective wavelengths of these filters are practically
identical to those of the corresponding Johnson bandpasses: 3597\AA, 4405\AA,
5495\AA, 6993\AA, and 9009\AA, respectively.  The total integration time for
RCrA, PV\,1, and Paley\,3 is about 20$\>$--$\>$30~min per field in each of
the $BVRI$ bands, and about 50~min in the $U$ band where the instrumental
efficiency is very low.  The integration times for the Paley\,1 field are
about a factor of~2 longer than in the corresponding bands for other
three~fields.  For photometric calibration purposes, the Landolt Selected
Area SA\,92 \markcite{landolt92}(Landolt 1992) containing several standard
stars was observed in $UBVRI$ twice during the first night.  Short twilight
flat exposures in all five~bands were obtained at the beginning and end of
each night; several series of bias frames were obtained during the daytime.
The FWHM of the stellar images ranges from $5\farcs1$--$5\farcs7$
(2.5$\>$--$\>$2.8~pixels), the result of atmospheric seeing, imperfect focus,
and coarse pixel scale.  The coordinates and observational parameters of the
clouds are listed in Table~\ref{coord_obs_param_tbl}.

\begin{deluxetable}{l c c r c c c c c c}
\tablewidth{0pt}
\tablecaption{Cloud Coordinates and Observational Parameters \label{coord_obs_param_tbl}}
\tablehead{
\colhead{(1)} & 
\colhead{(2)} & 
\colhead{(3)} &
\colhead{(4)} & 
\colhead{(5)} & 
\colhead{(6)} & 
\colhead{(7)} & 
\colhead{(8)} & 
\colhead{(9)} & 
\colhead{(10)}\\
\colhead{Name} & 
\colhead{$\alpha_{2000}$} &
\colhead{$\delta_{2000}$} & 
\colhead{$l$} &
\colhead{$b$} & 
\colhead{$t_{\rm int}(U)$} &
\colhead{$t_{\rm int}(B)$} & 
\colhead{$t_{\rm int}(V)$} & 
\colhead{$t_{\rm int}(R)$} & 
\colhead{$t_{\rm int}(I)$}\\
 & 
\colhead{($^{\rm h}$ $^{\rm m}$ $^{\rm s}$)} &
\colhead{($^\circ~^{\min}~^{\sec}$)} & 
\colhead{($^\circ$)} & 
\colhead{($^\circ$)} &
\colhead{(min)} & 
\colhead{(min)} &
\colhead{(min)} & 
\colhead{(min)} &
\colhead{(min)}\\
}
\startdata
RCrA    & 18 56 04.1 & $-$37 18 05 & 359.2 & $-$16.9 & 47 & 27 & 20 & 20 & 27 \\
PV\,1   & 22 06 28.4 & $-$03 32 33 &  56.5 & $-$44.1 & 53 & 27 & 20 & 20 & 27 \\
Paley\,1& 02 38 24.6 & $-$29 43 36 & 225.6 & $-$66.4 & 93 & 53 & 47 & 40 & 47 \\
Paley\,3& 23 48 53.8 & $-$71 47 12 & 309.7 & $-$44.5 & 53 & 33 & 20 & 20 & 33 \\
\tablecomments{\\
Col.~6$\>$--$\>$10: Total integration time in minutes in
each of the $U$, $B$, $V$, $R$, and $I$ bands.}
\enddata
\end{deluxetable}

\subsection{Data Reduction}
\label{datared}

As a first step, all the images (object, photometric calibration, and
flat-field exposures) are overscan- and bias-subtracted and trimmed.  The
twilight flats and selected disregistered (dark sky) object exposures of all
four~cirrus clouds are combined into a master flat-field image for each of
the $UBVRI$ bands; the master flat-field image is applied to all object and
photometric calibration exposures.  After flat fielding, image defects
(e.g.,~hot pixels, charge traps, and bad columns) are replaced by the median
value of the pixels surrounding the defects. 

A set of relatively bright and isolated stars are selected in each of the
four~cirrus fields, and these are used to determine the linear shift in
($x$,$\,y$) between each of the images and a particular astrometric reference
image.  After registering the images approximately, the IRAF routine {\sc
geomap} is applied to these same bright stars in order to determine higher
order coordinate transformations (translation, rotation, magnification,
distortion) for every image with respect to the reference image; the IRAF
routine {\sc geotran} is then used to accurately align all the images using
the coordinate transformations determined by {\sc geomap}.

The median background sky level for each object exposure is subtracted from
the image.  The registered, sky-subtracted images of a given cirrus cloud in
a given band are then combined into a median-averaged image.  In the case of
the PV\,1 field, the translational offsets between various exposures are
quite large (of the order of the field of view of an individual CCD image);
the final combined image in each band for PV\,1 is the union of all
the images in that band.  As a result, the effective exposure time, and hence
the r.m.s.\ sky noise level, varies with position across the final image of
the PV\,1 field.  For the other three~cirrus fields, only the intersection
area of all exposures is used, and the noisy (partial overlap) edges of the
combined images are discarded.

The final, combined $V$-band images of the four~cirrus cloud fields are shown
in Figure~\ref{v_images}.  The images contain several very bright, saturated
stars with extensive and complicated charge bleed patterns; the exact number
of saturated stars in the image depends on the Galactic latitude of the field
(most numerous in RCrA) and on wavelength (many more in $R$- and $I$-band
images than in $UBV$).  Instead of attempting to model (and subtract) the
intensity distribution of saturated stars, the affected regions in their
vicinity are eliminated from further consideration, since detection and
photometry of faint stars is unreliable in these regions.  This is done by
finding the positions of stars for which the central pixel value is above the
saturation limit ($\sim60,000$~ADU or $1.26\times10^5$~electrons) and
determining the number of saturated pixels at the center of every saturated
star.  An empirical relation is then defined between the number of saturated
pixels at the center of a stellar image and the size of the associated region
to be eliminated; this relation is used to mask a circular area of
appropriate radius around every saturated star. 


\subsection{Stellar Photometry}
\label{stellphot}

Lists of object positions are generated independently for each of the
five~bands and for each of the four~cirrus fields using the peak finding
algorithm {\sc find} of the stellar photometry program {\sc daophot}
\markcite{stetson87}\markcite{stetson92}(Stetson 1987, 1992).  An object
detection threshold of $3.5\sigma$ is used, where $\sigma$ is the sky noise;
in the case of the PV\,1 field, the variation of sky noise level across the
image (due to non-uniform effective exposure time) is taken into account.

Most of the detected objects are Galactic stars; a small fraction are distant
field galaxies.  Even at high Galactic latitudes
($\vert{b}\vert\gtrsim30^\circ$), the surface density of stars exceeds that
of field galaxies for $V\lesssim20$ \markcite{kron80}(cf.~Kron 1980), roughly
the apparent magnitude limit of our study (see Sec.~\ref{complete} and
Figs.~\ref{wolf_u_all} and~\ref{wolf_bvri_rcra}).  Given the relatively
coarse angular resolution of our CCD data and the fact that the majority of
the galaxies in the sample are expected to be close to the
detection/completeness limit, no attempt has been made to exclude galaxies
using morphological star-galaxy separation.  In fact, field galaxies are just
as useful as distant stars for studying the optical depth of a foreground
cirrus cloud.  In the rest of this paper, we loosely use the term ``stars''
to refer to the sample of detected (mostly stellar) objects.

An iterative procedure, based on a combination of several {\sc daophot}
tasks, is applied to the {\sc daophot}/{\sc find} star list in order to build
an empirical point spread function template and to fit the template to every
star in the list.  Aperture photometry is carried out for each star in turn,
with the results of the point spread function fit used to subtract off the
light of the neighbors of the star in question.  This combination of point
spread function fitting and aperture photometry techniques is ideal for
photometry of undersampled stellar images in crowded fields
\markcite{guha96}(Guhathakurta \etal\ 1996).  The aperture magnitudes,
calculated over small (2~pixel radius) apertures to avoid neighbor
contamination, are corrected to total magnitudes using standard
``curve-of-growth'' corrections.  The curve of growth is determined
independently for each band and for each cirrus field, based on the intensity
profiles of a set of isolated bright stars.  The lists of star positions and
corresponding total instrumental magnitudes are then combined into a
position-matched $UBVRI$ list for each cirrus field.  Total instrumental
magnitudes are also determined for photometric standard stars in the SA\,92
calibration field using a similar curve-of-growth method. 

As the SA\,92 calibration field was only observed on the first (photometric)
night, a bootstrap magnitude zeropoint adjustment is determined for each
field and for each band: the airmass-corrected instrumental magnitudes of a
sample of secondary standard stars in a single image obtained during that
first night are compared to those measured on the median-averaged image.
This zeropoint adjustment is quite small, ranging from
0.01$\>$--$\>$0.09~mag.  One of the four~fields (Paley\,3) was not observed
during the first night and thus could not be bootstrap corrected.  However,
the resulting error in the overall magnitude scale for this field is likely
to be small ($<0.1$~mag) and thus unimportant for the analysis in this paper.

The IRAF routine {\sc fitparam} is used to derive the zeropoints and color-
and airmass-coefficients of the transformation equations for converting
instrumental magnitudes to calibrated Johnson-Kron-Cousins $UBVRI$
magnitudes, based on measurements of the total instrumental magnitudes of
photometric standard stars in the SA\,92 field.  The instrumental stellar
magnitudes in the four cirrus fields are converted to calibrated $UBVRI$
magnitudes using the routine {\sc invertfit}, adopting an airmass term that
is the average of all the individual exposures for that band in each field.
The transformation equations from $(UBVRI)_{\rm inst}\rightarrow(UBVRI)_{\rm
calib}$ contain color terms based on $U-B$, $B-V$, $B-V$, $V-R$, and $R-I$,
respectively.  The corresponding color term coefficients are $-$0.034,
$+$0.165, $-$0.073, $+$0.027, and $-$0.013.  Not all stars are detected in
all bands; the combination of $BVI$ detections is of particular interest to
us (see Sec.~\ref{colorcolor}), including a small subset of stars which
happen to be undetected in the $R$ band.  To estimate the color term in the
$I_{\rm inst}\rightarrow{I}_{\rm calib}$ relation for these $R$-band
non-detections, their calibrated $R-I$ color is estimated using the empirical
correlation between $(B-V)_{\rm calib}$ and $(R-I)_{\rm calib}$ that is seen
for stars detected in all bands.  This technique is accurate enough for our
purposes since the $I$-band color term coefficient is very small.

For the star count analysis described in Sec.~\ref{counts}, star lists are
also constructed on the basis of detections in a single band (e.g.,~$U$ band
only, $B$ band only, etc.).  These single-band detections are calibrated
using an {\it average\/} color term in the instrumental $\rightarrow$ Johnson
conversion relation; for example: $U_{\rm calib}=U_{\rm inst}+\langle{U_{\rm
calib}-U_{\rm inst}}\rangle$, where the average is computed over all stars
for which a proper transformation is possible.  Thus the calibrated $UBVRI$
magnitudes in the single band lists are only approximations to true
calibrated magnitudes with proper color terms.  This approximation however
should be quite good as the color term coefficients are generally small. 

\subsection{Completeness}
\label{complete}

The completeness of the single-band star lists as function of magnitude,
$N_{\rm obs}/N_{\rm true}$, is modeled as a double exponential:

\begin{eqnarray}
N_{\rm obs}/N_{\rm true}~=~0.5\left[2-\exp((m-m_{50})/\Delta m)\right] 
\end{eqnarray}
\begin{center}
$m{\leq}m_{50}$\\
\end{center}
\begin{eqnarray}
N_{\rm obs}/N_{\rm true}~=~0.5\left[\exp(-(m-m_{50})/\Delta m)\right]
\end{eqnarray}
\begin{center}
$m>m_{50}$\\
\end{center}

\noindent
The underlying star count versus apparent magnitude relation, $N_{\rm true}$,
is assumed to be a power law in the vicinity of the 50\% completeness point,
$m_{50}$.  The values of the two free parameters $m_{50}$ and $\Delta m$
(sharpness with which incompleteness sets in) are then determined by fitting
to the shape of the observed count distribution, $N_{\rm obs}$.  At
magnitudes brighter than the completeness limit, the observed counts are a
good fit to the Bahcall-Soneira model \markcite{bahcall80}(Bahcall \& Soneira
1980; see Sec.~\ref{iasgmodel} and Fig.~\ref{wolf_u_all}), which predicts
$N_{\rm true}$ to be roughly a power law over the magnitude range of
interest.  The typical value of $\Delta{m}$ is about 0.4, with a range from
0.3 to 0.6, while the 50\% completeness values range from $I_{50}\approx18$
to $U_{50}\approx21$ (as indicated in Figs.~\ref{wolf_u_all}
and~\ref{wolf_bvri_rcra}).

\section{Far Infrared Data}
\label{irdata}

The {\it IRAS\/} 100\micron\ (ISSA) maps provided by the Infrared Processing
and Analysis Center are used to divide each optical image into three~regions
on the basis of 100\micron\ surface brightness, $F_{100\mu\rm{m}}$.  The mean
brightness, $\langle{F}_{100\mu\rm{m}}\rangle_{\rm bg}$, and pixel-to-pixel
r.m.s.\ variation, $\sigma_{100\mu\rm{m}}$, of the 100\micron\ background is
computed in regions without visible optical cirrus.  An ``off-cirrus'' region
is defined in which\\ 
$F_{100\mu\rm{m}}<\langle{F}_{100\mu\rm{m}}\rangle_{\rm
bg} + 5\sigma_{100\mu\rm{m}}$,\\ 
along with two ``on-cirrus'' regions:\\
$\langle{F}_{100\mu\rm{m}}\rangle_{\rm
bg} + 5\sigma_{100\mu\rm{m}}<F_{100\mu\rm{m}}<
\langle{F}_{100\mu\rm{m}}\rangle_{\rm bg}+11\sigma_{100\mu\rm{m}}$ (``on1'')\\
and\\
$F_{100\mu\rm{m}}>\langle{F}_{100\mu\rm{m}}
\rangle_{\rm bg}+11\sigma_{100\mu\rm{m}}$ (``on2'').\\
Each star in the matched $UBVRI$ list and single-band lists is then flagged
according to the region in which it falls.  The background 100\micron\ flux,
r.m.s.\ variation, and the areas of the ``off'', ``on1'', and ``on2'' regions
for each cirrus field are listed in Table~\ref{prop_clouds_tbl}.  These areas
are indicated on the $V$-band images of the four~clouds shown in
Figure~\ref{v_images}.

As discussed in Sec.~\ref{discuss} below, the optical depth of a cirrus cloud
can be estimated both from its effect on the counts/colors of background
stars (extinction optical depth) and from its far infrared brightness and
color temperature (emission optical depth).  The far infrared emission-based
estimate is derived from full sky 100\micron\ maps provided by
\markcite{schlegel98}SFD.  These maps are a reprocessed composite of the {\it
COsmic Background Explorer\/}/{\it Diffuse InfraRed Background Experiment\/}
({\it COBE\/}/{\it DIRBE\/}) and {\it IRAS\/} (ISSA) maps, with zodiacal
foreground and confirmed point sources removed.  The procedure combines the
relatively high angular resolution of {\it IRAS\/} 100\micron\ data ($3'$)
with the superior photometric calibration of {\it DIRBE\/} data, and is
estimated to be twice as accurate as the older \markcite{burstein82}Burstein
\& Heiles (1982) reddening maps in regions of low to moderate reddening.  In
addition to 100\micron\ flux maps with improved photometric calibration,
\markcite{schlegel98}SFD compute $F_{100\mu\rm{m}}^{\rm corr}$ over the full
sky, the equivalent 100\micron\ flux corrected for variations in mean dust
temperature from one line of sight to another.  The amount of reddening,
$E(B-V)$, is calculated by normalizing the amplitude of reddening per unit of
corrected 100\micron\ flux, based on the observed reddening towards brightest
cluster galaxies and elliptical galaxies.  The $E(B-V)$ values of
\markcite{schlegel98}SFD are scaled by $R_V\equiv{A_V}/E(B-V)=3.1$ to derive
differential on-cloud $A_V$ values (``on1''$-$``off'', ``on2''$-$``off'') for
each of the four~cirrus clouds in this study (Table~\ref{av_comp_tbl} and
Fig.~\ref{av_comp}).

\markcite{schlegel98}SFD present all-sky maps of the dust temperature,
averaged along the line of sight, derived from the ratio of {\it DIRBE\/}
100\micron\ to 240\micron\ flux measurements.  The effective angular
resolution of the temperature measurements is limited to $\sim1^\circ$ by the
resolution of the {\it DIRBE\/} far infrared data.  The mean dust temperature
of each of the four~cirrus clouds in this study (averaged over ``on1'' and
``on2'' regions) is listed in Table~\ref{prop_clouds_tbl}.  The implications
of cloud-to-cloud variations in dust temperature will be discussed in
Sec.~\ref{avcomp}.

\markcite{schlegel98}SFD assume a standard extinction law with $R_V=3.1$ in
deriving $E(B-V)$ values which should be correct on average over the full
sky.  The lines of sight towards the four~clouds in this study, however, may
have $R_V$ values that are different from this standard value (see
Sec.~\ref{extcurves} and~\ref{cumcolor}).  For a line of sight with a
non-standard extinction law, the \markcite{schlegel98}SFD value of $E(B-V)$
will be inaccurate, but the estimate of $A_V$ should be valid.
\markcite{schlegel98}SFD's estimate is effectively based on
$\tau_{100\mu\rm{m}}$ (correcting the 100\micron\ flux to the average dust
temperature).  The variation in extinction law from one line of sight to
another appears to be mostly in terms of the relative amount of extinction at
wavelengths at or shortwards of the $B$ band while the portion redward of the
$V$ band is invariant in shape
\markcite{cardelli89}\markcite{martin90}(Cardelli, Clayton, \& Mathis 1989,
hereafter CCM; Martin \& Whittet 1990).  In other words, our estimate of
$A_V$ is valid under the assumption that extinction laws along different
lines of sight, normalized by $\tau_{100\mu\rm{m}}$, are similar for
$\lambda\gtrsim\lambda_V$ and diverge only at shorter wavelengths.

\begin{deluxetable}{l c c c c c c}
\tablewidth{0pt}
\tablecaption{Properties of Cirrus Clouds \label{prop_clouds_tbl}}
\tablehead{
\colhead{(1)} & 
\colhead{(2)} & 
\colhead{(3)} &
\colhead{(4)} & 
\colhead{(5)} & 
\colhead{(6)} & 
\colhead{(7)}\\
\colhead{Name} & 
\colhead{$\langle{F}_{100\mu\rm{m}}\rangle_{\rm bg}$} & 
\colhead{$\sigma_{100\mu\rm{m}}$} & 
\colhead{Area$_{\rm off}$} & 
\colhead{Area$_{\rm on1}$} & 
\colhead{Area$_{\rm on2}$} & 
\colhead{$T_{\rm on}$}\\ 
 & 
\colhead{(MJy\,sr$^{-1}$)} & 
\colhead{(MJy\,sr$^{-1}$)} & 
\colhead{(deg$^{2}$)} & 
\colhead{(deg$^{2}$)} &
\colhead{(deg$^{2}$)} & 
\colhead{(K)}
}
\startdata
RCrA     & 9.3  & 0.6  & 0.61 & 0.21 & 0.15 & 17.8 \\
PV\,1    & 5.6  & 0.3  & 0.85 & 0.43 & 0.49 & 17.5 \\
Paley\,1 & ~0.19 & ~0.05 & 0.47 & 0.09 & 0.32 & 17.7 \\
Paley\,3 & 1.0  & 0.1  & 0.83 & 0.18 & 0.16 & 18.0 \\
\tablecomments{\\
\phantom{Col}Col. 2: Mean background 100\micron\ flux in regions without
  visible cirrus.\\
\phantom{Col}Col. 3: Pixel-to-pixel r.m.s.\ variation in background
  100\micron\ flux.\\
\phantom{Col}Col. 4--6: Area of ``off'', ``on1'', and ``on2'' regions
  ($V$-band), which refer to regions with fluxes that are $<5\sigma$,
  $5\sigma$--$11\sigma$, and $>11\sigma$ above the 100\micron\ background.
  Note, these areas are slightly different across $UBVRI$ because of
  differences in the area lost to saturated stars.\\
\phantom{Col}Col. 7: Mean dust temperature in on-cloud region from {\it
  DIRBE\/} far infrared measurements \markcite{schlegel98}(Schlegel et~al.\
  1998).}\\
\enddata
\end{deluxetable}

\section{Star Counts}
\label{counts}

\subsection{Analysis of Cumulative Star Counts}
\label{cumcounts}

Single-band star lists are used to derive star counts in each band and for
each region (``off'', ``on1'', and ``on2'') of each field.  The single-band
star lists are preferable to the matched $UBVRI$ lists for this purpose, as
they are based on clean, well-defined selection functions and contain more
stars by virtue of a higher degree of completeness at the faint end.
Figure~\ref{wolf_u_all} shows cumulative star counts versus apparent
magnitude, also called Wolf diagrams \markcite{wolf23}(Wolf 1923), for the
four~cirrus fields in the $U$ band; Figure~\ref{wolf_bvri_rcra} shows Wolf
diagrams for RCrA in $BVRI$.  As these diagrams are based on single-band
detections, the photometric conversion to the Johnson magnitude system is
only approximate (Sec.~\ref{optdata}).  The ``on1'' and ``on2'' distributions
shown in Figures~\ref{wolf_u_all} and~\ref{wolf_bvri_rcra} have been
normalized to match the area of the ``off'' region distributions.  Areas
around bright saturated stars (where the detection and photometry of faint
stars is unreliable) are excluded from the analysis.
Table~\ref{prop_clouds_tbl} lists the usable area of the $V$-band image,
properly accounting for the masked (excluded) regions around saturated stars.
These masked regions are slightly different in the different bands for a
given cirrus field; since saturation tends to be most severe at the longest
wavelengths, the $R$- and $I$-band images have typically smaller usable areas
than the $U$-band images.  The saturated stars themselves are included in the
star count analysis by adding the number of saturated stars in each field and
each band to the bright end of the cumulative distributions.  The total star
count ranges from $\sim24,000$ in $U$ to $\sim51,000$ in $V$ for RCrA, and
from $\sim2,000$ in $U$ to $\sim7,000$ in $V$ for Paley\,1, the fields with
the highest and lowest star densities, respectively, in our sample. 



A typical cumulative star count relation roughly resembles a power law, but
has slight curvature such that the slope,
$S_\lambda\equiv{d}\/\log[N(<m_\lambda)]/d\/m_\lambda$, gets shallower
towards fainter apparent magnitudes (Figs.~\ref{wolf_u_all}
and~\ref{wolf_bvri_rcra}).  The relations are uncertain at the bright end due
to the small number of stars; the abrupt flattening at the faint end marks
the onset of incompleteness.  There is a range of slopes from field to field
and a characteristic trend in $S_\lambda$ versus $UBVRI$, with the slopes
steepening towards both shorter and longer wavelengths relative to the $V$
band.  This is likely the result of two~competing effects.  On the one hand,
intrinsically faint (lower main sequence) stars tend to have redder colors
than intrinsically luminous stars (turnoff/upper main sequence), causing the
$I$-band slope to be steeper than the $V$-band slope.  On the other hand, the
``depth'' of the sample, in terms of the typical distance of stars at the
magnitude limit in each band, decreases from $V$ towards $U$.  The $U$-band
slope is close to the Euclidean value of~0.6 expected for a constant star
density, while the mean distance of the $V$-band sample is large enough for
the fall-off in Galactic star density along the line of sight to be
noticeable.  The star count slope is relevant to the discussion of the
effects of clumpiness (Sec.~\ref{clumpy}).

It is evident from Figures~\ref{wolf_u_all} and~\ref{wolf_bvri_rcra} that
dust extinction in the intervening cirrus cloud causes each on-cloud star
count relation to be shifted towards fainter magnitudes (i.e.,~to the right)
with respect to the corresponding off-cloud relation.  The magnitude of these
shifts decreases from $U$ to $I$ as expected on the basis of the normal
interstellar dust extinction law.  Moreover, the shift between the ``on2''
(the region with highest 100\micron\ flux) and ``off'' star count curves is
consistently larger than the shift between ``on1'' and ``off'' curves across
the $UBVRI$ bands, at least in the two~fields (RCrA and PV\,1) for which the
shifts are measured reliably for both ``on2'' and ``on1'' regions.

\subsection{Star Counts-Based Extinction Curves}
\label{extcurves}

The magnitude shift between Wolf diagrams on and off an obscuring cloud is a
direct measure of the effective dust extinction, $A_\lambda^{\rm counts}$, at
least at faint apparent magnitudes where the contribution of low luminosity
foreground stars is unimportant and the on and off curves run parallel to
each other.  As explained in Sec.~\ref{clumpy} below, the $A_\lambda^{\rm
counts}$ value derived from the magnitude shift in a Wolf diagram yields a
measure of the effective ``extinction'' that depends on the degree of
clumpiness of the dust distribution (even for a given mean dust optical depth
and a given set of grain optical properties).

Wolf diagram shifts are plotted in Figure~\ref{wolf_shifts} as a function of
apparent magnitude in the $UBVRI$ bands for all four~cirrus cloud fields.
The median value of $A_\lambda^{\rm counts}$ is computed over an intermediate
range of apparent magnitudes, avoiding excessive Poisson noise at the bright
end and the divergence of the Wolf diagram shift at the faint end caused by
the flattening of the cumulative star counts due to incompleteness.  In the
cases of the ``on1'' and ``on2'' regions of the RCrA and PV\,1 fields, the
Wolf diagram shifts are sufficiently large for the extinction to be
measurable in all five~bands.  For Paley\,1 and Paley\,3 on the other hand,
the extinction is lower and the measurement errors larger (see discussion of
errors below) so that a reliable measurement is only possible at the shortest
wavelengths and that too for only the ``on2'' regions ($U$ band for Paley\,1,
$U$ and $B$ bands for Paley\,3); for these two fields the $A_U$ and $A_B$
values are scaled to $A_V$ assuming a specific form of the extinction law,
either a standard law with $R_V=3.1$ or an $R_V=1.7$ law
\markcite{cardelli89}(CCM).


Figure~\ref{wolf_shift_err} provides an illustration of the typical errors in
the determination of the Wolf diagram apparent magnitude shift
$A_\lambda^{\rm counts}$.  The errors in $A_\lambda^{\rm counts}$ are derived
from the statistical uncertainties in the star counts, projected onto the
magnitude axis (abscissa) using the slope of the
log[$N$($<m_\lambda$)]-vs-$m_\lambda$ relation (roughly a power law over the
apparent magnitude range of interest), combined with the photometric
uncertainties.  The errors are of the order of~0.2$\>$--$\>$0.3~mag for RCrA,
the field with the largest number of stars.  Poisson error dominates
throughout most of the plotted apparent magnitude range; the contribution of
the photometric error becomes noticeable beyond $V\sim17\>$--$\>$18 in RCrA
(somewhat fainter in Paley\,3).  Increasing photometric error, in conjunction
with the flattening of the cumulative star count relations, causes the
flaring of the $A_\lambda^{\rm counts}$ error bands at the faint end.


The effective extinction curves for the ``on1'' and ``on2'' regions of RCrA
and PV\,1 derived from star counts are shown as open squares and triangles,
respectively, in Figure~\ref{rcra_pv1_ext_curves}.  The $A_\lambda^{\rm
counts}$ estimates are uncertain for the Paley\,1 and Paley\,3 fields (all
but the ``on2'' estimates at the shortest wavelengths) so extinction curves
are not presented for these two~clouds.  The lines are analytic
approximations to the extinction law provided by \markcite{cardelli89}CCM for
different $R_V$ values.  The RCrA and PV\,1 extinction curves are reasonable
approximations to the standard extinction law with $R_V=3.1$ (solid line).

Looking more closely, the PV\,1 ``on2'' extinction curve rises more steeply
towards short wavelengths ($U$ band) than the standard extinction law,
suggesting a lower value of $R_V$.  A minimum of the $\chi^2$ statistic is
obtained for $R_V=1.7$ (dashed line).  For RCrA on the other hand, the
increase of $A_\lambda^{\rm counts}$ with decreasing wavelength is somewhat
more gradual than for the standard extinction curve, with $A_I^{\rm counts}$
being particularly discrepant.  While this would appear to favor high $R_V$
values in that the $\chi^2$ statistic decreases with increasing $R_V$, there
is no clear $\chi^2$ minimum over the plausible range of parameter space
($R_V<7$).  Alternatively, a good fit to the RCrA $A_\lambda^{\rm counts}$
measurements is obtained by assuming a non-uniform dust density (dotted
line; Sec.~\ref{clumpy}) with $R_V=3.1$ or even with $R_V=1.7$, the lower
$R_V$ value being preferable in light of the color excess data for this cloud
(see Sec.~\ref{cumcolor}).  We will return to a discussion of the
counts-based extinction curves in Sec.~\ref{clumpy}.

The weighted average of the $A_\lambda^{\rm counts}$ measurements in $UBVRI$
is computed for RCrA and PV\,1, after the measurement in each band is scaled
to the equivalent $A_V$ estimate using the analytic form of the extinction
law for a specific $R_V$  (Table~\ref{av_comp_tbl} and Fig.~\ref{av_comp}).
While the choice of $R_V$ affects the quality of the fit to the
five~$A_\lambda^{\rm counts}$ measurements, the weighted average $A_V^{\rm
counts}$ is practically independent of $R_V$, and tends to be similar to the
measured value of $A_\lambda^{\rm counts}$ in the $V$ band.  In other words,
changing $R_V$ results in a steep or shallow extinction curve, but the
best-fit curves intersect near the middle of the range of available data
points which happens to be the $V$ band.


\subsection{Star Count Models}
\label{iasgmodel}

Wolf diagrams have traditionally been used to determine both the extinction
through and the distance to dust clouds.  Distance determination, in
particular, requires knowledge of the stellar luminosity function and the
spatial density distribution of stars along the line of sight.  Based on this
information, theoretical Wolf diagrams can be constructed for various values
of the dust extinction and distance of an obscuring cloud.  These diagrams
can then be directly compared to the star count data. 

Theoretical Wolf diagrams are generated for the four~cirrus clouds in this
study using the IASG Galaxy Model program
\markcite{bahcall87}\markcite{ratna89}\markcite{caser89}(Bahcall, Casertano,
\& Ratnatunga 1987; Ratnatunga, Bahcall, \& Casertano 1989; Casertano,
Ratnatunga, \& Bahcall 1989), which is based on the Bahcall-Soneira star
count model of the Galaxy \markcite{bahcall80}\markcite{bahcall86}(Bahcall \&
Soneira 1980; Bahcall 1986).  The IASG program yields the total projected
density of stars and their distribution as a function of apparent $V$
magnitude, $B-V$ color, and distance modulus, for any specified direction.
A Monte-Carlo realization of these distributions is carried out in order to
construct a stellar sample.  Next, $U-B$ colors are computed using the
\markcite{bert94}Bertelli \etal\ (1994) stellar evolution models, averaging
over all available ages per metallicity, for four~different metallicities,
$Z=0.0004$, $Z=0.008$, $Z=0.05$ and $Z=0.02$.  The product is a list of stars
with $UBV$ magnitudes and distance moduli for each metallicity from which
cumulative star counts are computed appropriate for the off-cloud region.
The dust cloud is assumed to be (geometrically) thin; it produces an apparent
offset of $A_\lambda$ for all stars located at distances greater than that
assigned to the dust cloud.  The resulting list is used to construct on-cloud
model star count relations.  As the model results are similar for the
four~different metallicities, only curves computed with $Z=0.05$ are shown.

The slopes of the model cumulative star count relations, plotted as dotted
blue lines in Figures~\ref{wolf_u_all} and~\ref{wolf_bvri_rcra}, agree well
with the observed slopes, with the possible exception of PV\,1.  An offset
has been applied to the apparent magnitude scale in the Bahcall-Soneira model
in order to match the observed counts in the ``off'' region: $+$0.2, $+$1.5,
$-$0.2, and $-$0.5~mag for RCrA, PV\,1, Paley\,3 and Paley\,1, respectively.
Positive corrections to the apparent magnitude may indicate that the overall
extinction integrated over the line of sight is not quite zero even in the
off-cloud regions.  There may, however, also be small systematic errors in
the model (e.g.,~in the apparent magnitude scale, or in the normalization of
the stellar density or luminosity function) that cause it to deviate from the
observed counts.  Note that since the curves (model and observed counts) are
good approximations to single power laws, a horizontal shift is roughly
equivalent to a vertical shift [in the log($N$) direction].

\subsection{Cloud Distance Determination}
\label{distance}

Model on-cloud and off-cloud Wolf diagrams are constructed for the RCrA and
PV\,1 fields using IASG star count predictions coupled with a geometrically
thin dust slab located at various line-of-sight distances.  These model
curves are shown as dotted blue lines in the upper panel of
Figures~\ref{wolf_u_all} and~\ref{wolf_bvri_rcra} for a cloud distance of
550~pc and an optical depth of $A_U=0.9$ (matching the ``on2'' measurements
of $A_U^{\rm counts}$ in the two~clouds), which scales to $A_V=0.6$ on the
basis of a standard extinction law.  The on-cloud model curves for RCrA shown
in the center and bottom panels of Figure~\ref{dist_method} are for cloud
distances of 400~pc and 800~pc, respectively, for optical depths of $A_U=0.9$
(``on2''; left) and $A_U=0.3$ (``on1''; right).

Only RCrA and PV\,1, the fields with the lowest latitiude and largest area,
respectively, contain enough bright stars ($U\sim13\>$--$\>$15; see
discussion below) and high enough extinction for us to obtain reliable
constraints on the distance using this method.  Figure~\ref{dist_method}
compares the RCrA ``on2'' and ``on1'' Wolf diagrams (upper left and upper
right panels, respectively) with their corresponding IASG model-based Wolf
diagrams for cloud distances of $d=0.4$~kpc (center panel) and $d=0.8$~kpc
(bottom panel), bracketing the likely distance of the RCrA cloud.  The best
estimate of the RCrA cloud distance is in the range 550$\>$--$\>$750~pc.  For
PV\,1, it is only possible to place an upper limit to the distance of about
1~kpc.  The distance determination is uncertain for PV\,1 due to its low
stellar surface density: the Poisson error in this field is 40\% larger than
in RCrA despite its $3.3\times$ larger area.


The IASG model Wolf diagrams shown in the center and bottom panels of
Figure~\ref{dist_method} are based on a very large number of synthetic stars
in order to minimize the effect of Poisson error.  A realistic measure of the
dispersion in cumulative star count relations is obtained by making several
Monte Carlo realizations of the on-cloud model, with each realization
containing the number of stars appropriate for the actual area of the RCrA
on-cloud (``on2'' or ``on1'') region being simulated.  The dispersion among
the realizations of a given model can be used to judge the significance with
which the model fits or is ruled out by the data; the $\pm1\sigma$ error bars
in the center left panel of Fig.~\ref{dist_method} indicate the dispersion
for the RCrA ``on2'' $d=0.4$~kpc model.  Averaging over the
$U=12\>$--$\>$15~mag range, 66\% of the $d=0.4$~kpc model realizations
lie below the RCrA ``on2'' star count curve, while 82\% of the $d=0.8$~kpc
model realizations lie above it (both models adopt $A_U=0.9$).  Comparing the
$A_U=0.3$ model to the RCrA ``on1'' star count curve, 54\% of the $d=0.4$~kpc
realizations lie below the data while 90\% of the $d=0.8$~kpc realizations
lie above the data.

The differences between the on-cloud and off-cloud model cumulative star
count curves can be understood in simple terms.  The transition between
majority-foreground and majority-background in the on-cloud star count
relation occurs at an apparent magnitude equal to that of a ``typical'' star
at the distance of the cloud.  While stars have a wide range of absolute
magnitudes, the characteristic absolute magnitude is defined by an abrupt
rise near the bright end of the stellar luminosity function.  The rise is at
the main sequence turnoff, $M_V^{\rm MSTO}\approx+3.5$, for any old stellar
population, and is particularly steep in the $U$ band ($M_U^{\rm
MSTO}\approx+4.0$).  Thus a cirrus cloud at a distance of 0.4~kpc (0.8~kpc),
corresponding to a distance modulus of $(m-M)_0=8.0$~mag (9.5~mag), is
expected to have a transition region at $U\sim12.0$ ($U\sim13.5$), in
agreement with the model curves shown in the center (bottom) panel of
Figure~\ref{dist_method}.

\section{Effect of Reddening on Stellar Colors}
\label{colors}

In this section, the distribution of stellar colors is compared between on-
and off-cloud regions for the purpose of exploring the reddening (selective
extinction) caused by dust.  In contrast to the star count analysis described
in the previous section, the color analysis described in this section makes
no use of the number or number density of stars in the various regions, and
only uses information about the normalized stellar color distribution in each
region (``off'', ``on1'', and ``on2'').  As discussed in Sec.~\ref{discuss}
below, star counts and color distributions provide complementary measures of
the optical depth of a cirrus cloud. 

\subsection{Cumulative Color Distributions and Color Excess Measurements}
\label{cumcolor}

A direct way to determine the amount of reddening caused by the cirrus cloud
is to measure the difference in the stellar colors between on- and off-cloud
regions.  This is done by constructing normalized cumulative color histograms
in $U-V$, $B-V$, $R-V$, and $I-V$ for the ``off'', ``on1'', and ``on2''
regions of the cirrus cloud fields.  The median horizontal (i.e.,~color)
difference between on-cloud and off-cloud histograms is defined to be the
color excess, $E(m_\lambda-V)\equiv{A}_\lambda-A_V$, for $\lambda=UBRI$.
Sample cumulative $U-V$ and $I-V$ color histograms are displayed in
Figure~\ref{cum_color} (solid and dashed lines, respectively) for the ``on2''
and ``off'' regions (thin and bold lines, respectively) of RCrA (upper panel)
and Paley\,3 (lower panel), with the horizontal line segment indicating the
median value of the offset in each case.  The color excess is well determined
for most fields and colors, with the exception of long wavelength
measurements for the low optical depth clouds, Paley\,1 and Paley\,3 [as
illustrated by the Paley\,3 $E(I-V)$ data in Fig.~\ref{cum_color}].

If a small fraction of the area of the ``on1'' or ``on2'' region of the cloud
happens to have a much larger optical depth than the rest, it is effectively
ignored by the use of the median in the color offset computation; the
significance of this aspect of our measurement technique is discussed further
in Sec.~\ref{clumpy}.  The median offset is also insensitive to dilution by
unreddened, bright, foreground stars, especially as these are greatly
outnumbered by background stars.


The median color excess, $E(m_\lambda-V)$ for $m_\lambda=UBRI$, for the
``on1'' and ``on2'' regions of the four~cirrus clouds is plotted as open
squares and triangles in Figure~\ref{color_excess}.  The dashed and solid
lines show the \markcite{cardelli89}CCM analytic approximation for the
Galactic interstellar reddening law for $R_V=3.1$ for ``on1'' and ``on2'',
respectively, while the dotted line shows an $R_V=1.7$ reddening law for
``on2''.  Each analytic curve is scaled by the appropriate weighted mean
value of the visual optical depth, $A_V^{\rm excess}$, derived from the
four~color excess measurements (listed in Table~\ref{av_comp_tbl} and shown
in Fig.~\ref{av_comp}).  As expected, $A_V^{\rm excess}$ is sensitive to the
choice of $R_V$; this is in contrast to the $A_V^{\rm counts}$ estimate
derived from the $A_\lambda^{\rm counts}$ measurements
(Sec.~\ref{extcurves}).


The measured color excesses tend to increase more steeply with increasing
$1/\lambda$ ($V\rightarrow{U}$) than the best-fit standard $R_V=3.1$
reddening law, and vary less strongly with $\lambda$ from $V\rightarrow{I}$
than the standard reddening law.  In all cases, the $R_V=1.7$ reddening law
shown in Figure~\ref{color_excess} appears to be a better match to the shape
of the observed $E(m_\lambda-V)$ versus $1/\lambda$ than the standard
reddening law.  Applying a $\chi^2$ test to the RCrA and PV\,1 color excess
data, a minimum of the $\chi^2$ statistic, albeit a relatively shallow one,
is found for $R_V$ values that are substantially lower than the standard
Galactic value of~3.1.  The Paley\,1 and Paley\,3 $\chi^2$ behavior also
favors a low value of $R_V$ ($\lesssim2$) but there is no clear minimum as
the Poisson error in these fields is larger than for RCrA or PV\,1.

Note, the $\chi^2$ test is only used in a relative way here.  The $\chi^2$
values are {\it not\/} used to decide if the \markcite{cardelli89}CCM
analytic form of the reddening law with a specific value of $R_V$ is a good
fit to the data in an absolute sense; the exact value of $\chi_{\rm min}^2$
is not particularly meaningful because: (1)~the $E(m_\lambda-V)$ error
estimates used in the $\chi^2$ test are only approximate representations of
the true (likely non-Gaussian) error distribution; and (2)~the
$R_V$-parametrized analytic formula used in the analysis is known to deviate
from the measured reddening law in certain lines of sight
\markcite{cardelli91}(Cardelli \& Clayton 1991).

\subsection{Color-Color Diagrams}
\label{colorcolor}

A second way to determine the cloud optical depth is through the use of
color-color diagrams.  This method is similar in principle to the color
excess method described above except that a pair of stellar colors [e.g.,~the
($U-B$,~$B-V$) pair or the ($B-V$,~$V-I$) pair] is analyzed jointly.  As
the intrinsic spread in stellar $V-R$ and $R-I$ colors is not very broad
relative to the photometric errors in these quantities, the $R$-band data are
excluded from this analysis and the longer $V-I$ color baseline is used
instead for a more accurate measurement of the reddening.

Figure~\ref{color_color} shows the $U-B$ versus $B-V$ and $V-I$ versus $B-V$
color-color diagrams (upper and lower panels, respectively) for RCrA and
PV\,1 (left and right panels, respectively).  The Paley\,1 and Paley\,3
fields contain too few stars and the reddening in these clouds is too low for
an analysis of their color-color distributions.  The black, red, and green
points indicate stars in the ``off'', ``on1'', and ``on2'' regions,
respectively, plotted in this order.  The blue lines show theoretical loci,
computed from the \markcite{bert94}Bertelli \etal\ (1994) stellar evolution
models by averaging all available ages per metallicity, for
three~metallicities: $Z=0.0004$, $Z=0.008$, and $Z=0.05$ (top to bottom).
The large spread of points is due to a combination of photometric errors,
intrinsic range of stellar colors (spread in metallicity and age), sample
contamination by background galaxies, spurious detections/matches, etc..


The arrows in Figure~\ref{color_color} indicate reddening vectors
\markcite{cardelli89}(CCM) corresponding to $A_V=2$~mag for a standard
extinction law with $R_V=3.1$ (red) and for an $R_V=1.7$ extinction law
(black).  The latter extinction law is a reasonable approximation to the
available star count and color excess data for the four~clouds.  The
progressive shift along the reddening vector of the ``on1'' and ``on2''
stellar color-color distributions with respect to the ``off'' distribution is
clearly visible in RCrA, and is somewhat less obvious in PV\,1.  The
principal direction along which the stellar distribution is extended is
almost parallel to the reddening vector, especially in the $V-I$ vs $B-V$
plot; even so, an accurate statistical measurement of the reddening is
rendered possible by the use of a well-matched off-cloud region and the large
numbers of stars in the sample (which enables accurate determination of the
centroid of the stellar distribution).

The amount of reddening in the color-color diagrams is estimated by measuring
the shifts between on- and off-cloud stellar distributions projected onto the
direction of the reddening vector.  The sample of stars is divided into a
series of contiguous slices parallel to the reddening vector each of width
0.1~mag.  The median offset along the length of the slice is computed between
the ``on1''/``on2'' and ``off'' distributions for each slice.  The final
visual optical depth estimate is a weighted average of the median shifts
derived from the different slices.  These optical depth estimates are denoted
$A_V^{UBV}$ and $A_V^{BVI}$ for the ($U-B$,~$B-V$) and ($V-I$,~$B-V$) plots,
respectively.  The offsets (in $A_V$ units) for the individual slices of the
RCrA $U-B$ vs $B-V$ diagram are illustrated in Figure~\ref{shift_ubv} for
``on2'' (solid line) and ``on1'' (dashed line).  The error bars indicate the
combined standard error in the mean centroids of the on- and off-cloud
stellar distributions.  The weighted mean of the individual slice offsets is
indicated by the horizontal dot-dashed line.

A test is carried out in order to assess the effect of incompleteness at the
faint end of the sample, or, more specifically, the effect of differential
incompleteness between on- and off-cloud samples which may introduce a bias
in the color offset measurement.  The ``off'' sample is artifically reddened
by an arbitrary amount corresponding to an optical depth of $A_V^{\rm sim}$.
This simulated reddened sample is then folded through the completeness
function in the three~bands of interest (i.e.,~$UBV$ or $BVI$).  The offset
of the on-cloud sample is measured with respect to the simulated reddened and
truncated sample.  The process is iterated while varying $A_V^{\rm sim}$
until there is a null offset between simulated and on-cloud samples.  The
value of $A_V^{\rm sim}$ that produces a null offset is found to be virtually
identical to the directly measured value of the visual optical depth
($A_V^{UBV}$ or $A_V^{BVI}$), so it may be concluded that differential
incompleteness effects are unimportant in the determination of $A_V$ from
color-color diagrams.


The calculation of $A_V^{UBV}$ and $A_V^{BVI}$ for the ``on1'' and ``on2''
regions of RCrA and PV\,1 is done two~ways, assuming a standard $R_V=3.1$
reddening law or the steeper $R_V=1.7$ reddening law.  The resulting $A_V$
estimates are listed in Table~\ref{av_comp_tbl} and shown in
Figure~\ref{av_comp}.  The second set of $A_V$ values are understandably
lower than the first: a lower $R_V$ produces a steeper extinction law in the
blue portion of the spectrum (i.e.,~stronger variation with $\lambda$), so
that a fixed (observed) color offset corresponds to a smaller $A_V$.  A
secondary effect is that the directions of the $R_V=3.1$ and $R_V=1.7$
reddening vectors are slightly different in the ($V-I$,~$B-V$) plane and this
leads to different projections of the stellar color-color offsets.

In contrast to the star count and color excess analysis described above
(Sec.~\ref{extcurves} and~\ref{cumcolor}), a specific form of the reddening
law had to be assumed {\it a~priori\/} in deriving $A_V$ from color-color
diagrams.  Given the similar orientations of the $R_V=3.1$ and $R_V=1.7$
reddening vectors (directions almost degenerate in $U-B$ vs $B-V$ plot), it
is impossible to determine {\it both\/} $R_V$ and $A_V$ independently from a
single color-color diagram, even for a field containing as many stars as
RCrA.  It would require a very large, uncontaminated sample of stars and
accurate photometry to exploit the subtle dependence of reddening vector
direction on $R_V$ in the ($V-I$,~$B-V$) plane.  Alternatively, the
combination of all four~available colors can discriminate between different
$R_V$ values (cf.~Sec.~\ref{cumcolor}).

\section{Discussion}
\label{discuss}

\subsection{Reddening Law in Diffuse Cirrus Clouds}
\label{redlaw}

Studies of the interstellar extinction law along various lines of sight in
the Galaxy show some evidence for a dependence of $R_V$ on environment.
Dense regions such as optically thick molecular clouds, of which the
$\rho$\,Oph dark cloud is a typical example \markcite{martin90}(Martin \&
Whittet 1990), tend to have relatively shallow extinction curves indicating
$R_V$ values larger than the global average of~3.1, while regions of low
density display extinction curves that rise steeply towards short wavelengths
indicating lower-than-average $R_V$ values (cf.~\markcite{mathis90}Mathis
1990), although some counter examples do exist.  The fact that all
four~diffuse clouds in our sample have reddening laws that favor lower $R_V$
values is in keeping with this trend.  Also consistent with the findings of
this paper is the observed steepening of the far-ultraviolet extinction curve
with increasing distance from the Galactic plane
\markcite{kiszkurno87}(Kiszkurno-Koziej \& Lequeux 1987).  On the face of it,
the RCrA data appear to yield conflicting results: a higher-than-average
$R_V$ value fits the star counts-based extinction curve (Sec.~\ref{counts})
while a lower-than-average value fits the color excess measurements.  As
explained in Sec.~\ref{extcurves} and~\ref{clumpy}, however, all of the RCrA
data are adequately explained by $R_V\lesssim2$ if one invokes a non-uniform
dust density. 

The $R_V$ values for the four~cirrus clouds in this study are unusually low,
although comparable values have been observed in a few instances.  A study of
about 200~OB stars by \markcite{whittet80}Whittet \& van Breda (1980) found a
minimum value of $R_V=2.4$, while the subsequent survey by
\markcite{fitz90}Fitzpatrick \& Massa (1990) found a minimum value of
$R_V=2.6$ for the line of sight towards HD\,204827.  More recently,
\markcite{larson96}Larson, Whittet, \& Gough (1996) reported a value of
$R_V=2.1$ for the high-latitude cloud DBB\,80 based on their study of the
early type star HD\,210121.

The observed variations in the shape of the interstellar extinction law from
one line of sight to another are predominantly at blue and ultraviolet
wavelengths while the red and infrared portion remains invariant
\markcite{martin90}(Martin \& Whittet 1990).  Variations in $R_V$, the ratio
of selective blue-minus-visual extinction to total extinction, are generally
thought to be a result of variations in the size distribution of dust grains:
the total visual optical depth is determined by the abundance of large
grains, while the extinction in the blue and ultraviolet portion is dominated
by smaller grains \markcite{cardelli89}\markcite{larson96}(cf.~CCM; Larson
\etal\ 1996).  Thus, a steep extinction curve is caused by a higher than
average proportion of small size particles in a dust cloud.  An abundance of
small particles may be explained in terms of the cirrus being in a UV-poor
and weak radiation environment in which grain destruction by photon heating
is inefficient.  An alternative explanation is that the growth, by
coagulation, of larger size particles is inhibited by the low overall density
of the cloud \markcite{larson96}(Larson \etal\ 1996).

\subsection{Effect of a Clumpy Medium}
\label{clumpy}

In this section, we investigate the effect of a clumpy cirrus cloud on the
counts and colors of background stars.  The technique of using normalized
color histograms to measure the reddening caused by cirrus
(Sec.~\ref{colors}) yields the average optical depth for the lines of sight
towards the specific set of stars used in the color analysis.  If there is
substantial non-uniformity in the dust column density within a given cloud
region (``on1'' or ``on2''), so much so that a highly-reddened subset of
background stars drops out of the matched $UBVRI$ sample, the color-based
method will underestimate the true area-weighted average optical depth.  Wolf
diagrams, on the other hand, are sensitive (in a statistical sense) to stars
that have dropped out of the on-cloud sample (Sec.~\ref{counts}).  In
addition to the ``horizontal'' shift to fainter apparent magnitudes of the
on-cloud cumulative star count curve due to extinction of stars in the
sample, any highly-attenuated undetected subset of stars will cause a
``vertical'' downward shift in the curve, which is equivalent to a shift to
the right since the star count relation closely resembles a power law.

The characterization of a clumpy medium is, in general, complicated and
beyond the scope of this paper.  Instead, the essential features of a
non-uniform dust cloud are illustrated with the help of a simple two-phase
model: a fractional area $f$ of the absorbing material is assumed to have a
uniform and relatively low optical depth of $A_V^{\rm sheet}$ while the
remaining fraction $(1-f)$ is covered by dense clumps with optical depth
$A_V^{\rm clump}$.  Exact calculations are carried out for the particular
case: $A_V^{\rm clump}\rightarrow\infty$.  Since the Wolf diagram shift is
statistical in nature, the following calculations involve no assumptions
about the morphology of the low and high extinction regions.  However,
considering that a large contiguous area with $A_V\rightarrow\infty$ would be
conspicuous in the CCD images, a low extinction sheet with numerous small
dense clumps seems a more likely scenario.

With the above assumptions, and given the specifics of the measurement method
used (Sec.~\ref{colors}), the color excess histograms and color-color
diagrams measure the extinction in the smooth, low optical depth part of the
cloud, $A_\lambda^{\rm sheet}$.  The Wolf diagram shift is greater than
$A_\lambda^{\rm sheet}$ but its exact value depends on the shape of the
cumulative star count function.  For the purposes of this calculation, the
off-cloud number count relation is assumed to be a power law with slope
$S_\lambda$: $N_{\rm off}(<m_\lambda)=C_\lambda10^{S_\lambda{m}_\lambda}$,
where $C_\lambda$ is a normalization constant.  The cumulative star count
function in the on-cloud (``on1'' or ``on2'') region is then a power law with
the same slope: $N_{\rm
on}(<m_\lambda)=C_\lambda\left[f\>10^{S_\lambda(m_\lambda-A_\lambda^{\rm
sheet})}+(1-f)10^{S_\lambda(m_\lambda-A_\lambda^{\rm clump})}\right]$.  Thus,
the Wolf diagram shift, $(A_\lambda^{\rm counts})_{\rm model}$, is directly
proportional to the ``vertical'' shift between the on- and off-cloud
relations at a fixed apparent magnitude $m_\lambda$.

\begin{eqnarray}
(A_\lambda^{\rm counts})_{\rm model} =
 \frac{1}{S_\lambda}\left[\log(N_{\rm off})-\log(N_{\rm on})\right]\\
 = \frac{-1}{S_\lambda}\log\left[f\>10^{-S_\lambda{A}_\lambda^{\rm
 sheet}} + (1-f)10^{-S_\lambda{A}_\lambda^{\rm clump}}\right]
\end{eqnarray}

\noindent
Under the assumption that $A_\lambda^{\rm clump}>\!\!>1$ over the wavelength
range of interest, this simplifies to:

\begin{eqnarray}
(A_\lambda^{\rm counts})_{\rm model} = \frac{-\log(f)}{S_\lambda} +
 A_\lambda^{\rm sheet}
\label{extlaw_eqn}
\end{eqnarray}

\noindent
where the first~term accounts for the undetected fraction of background stars
while the second~term is the extinction caused by the smooth sheet on the
rest of the stars.  The $V$-band version of the above equation may be
rewritten as:

\begin{eqnarray}
\log(f) = S_V\,\left[A_V^{\rm sheet}-(A_V^{\rm counts})_{\rm model}\right]
\label{logf_eqn}
\end{eqnarray}

\noindent
Normalizing the counts-based ``effective'' extinction (Eq.~\ref{extlaw_eqn})
by the $V$-band value and substituting for $f$ using Eq.~\ref{logf_eqn}, it
follows that:

\begin{eqnarray}
\left[\frac{A_\lambda^{\rm counts}}{A_V^{\rm counts}}\right]_{\rm model}
 = \frac{S_V}{S_\lambda} +
 \left[ r_\lambda - \frac{S_V}{S_\lambda} \right]\>
 \frac{A_V^{\rm sheet}}{(A_V^{\rm counts})_{\rm model}}
\label{normextlaw_eqn}
\end{eqnarray}

\noindent
where $r_\lambda\equiv(A_\lambda/A_V)$ depends only on the shape of the
extinction law.  The $UBVRI$ star count slopes vary somewhat from field to
field (see Sec.~\ref{cumcounts} for a discussion of the general trends in
$S_\lambda$).

A one-parameter family of normalized counts-based ``effective'' extinction
curves, computed by varying the ratio $A_V^{\rm sheet}/(A_V^{\rm
counts})_{\rm model}$ for a standard $R_V=3.1$ extinction law (CCM) and using
the observed star count slopes $S_\lambda$ in the RCrA field, is shown in
Figure~\ref{clumpy_ext_curves}.  An increasing degree of non-uniformity
(decreasing $f$) results in a flattening of the counts-based extinction
curve.  The effect of clumpiness is qualitatively similar for other
extinction laws as well (e.g.,~$R_V=1.7$).  Although Eq.~\ref{normextlaw_eqn}
depends only on the ratio $A_V^{\rm sheet}/A_V^{\rm counts}$ and not on the
value of $A_V^{\rm counts}$ itself, the derived value of the area covering
fraction $f$ does depend on $A_V^{\rm counts}$ (Eq.~\ref{logf_eqn}).
Moreover, the normalized counts-based extinction curves display a similar
behavior with increasing clumpiness even if the $A_V^{\rm
clump}\rightarrow\infty$ assumption is relaxed, although their exact shapes
in that case depend somewhat on the value of $A_V^{\rm counts}$.


There is an apparent discrepancy in the case of RCrA.  The Wolf diagram
shifts for this cloud, $A_\lambda^{\rm counts}$, increase less steeply with
decreasing wavelength than the standard $R_V=3.1$ extinction law, indicating
a larger $R_V$ value (Sec.~\ref{extcurves}).  On the other hand, the color
excesses for the same cloud are a steep function of wavelength and are
adequately described by the $R_V=1.7$ CCM reddening law
(Sec.~\ref{cumcolor}).  This discrepancy can be resolved if the dust column
density in RCrA is non-uniform.  For example, the $A_\lambda^{\rm counts}$
measurements can be fit by a combination of $R_V=1.7$ and a smooth sheet
fraction of $f=0.8$, which corresponds to $A_V^{\rm sheet}/A_V^{\rm
counts}=0.5$ and $A_V^{\rm sheet}=0.3$ (this translates to $f=0.7$ for
$A_V^{\rm counts}=1$ in Fig.~\ref{clumpy_ext_curves}).  The combinations
($R_V=3.1,~f=0.9$) and ($R_V\sim5,~f=1$) produce $A_\lambda^{\rm counts}$ vs
$\lambda$ functions that are practically identical in shape to the
($R_V=1.7,~f=0.8$) case and they all fit the RCrA measurements equally well
(see dotted line in Fig.~\ref{rcra_pv1_ext_curves}).  Considering the low
$R_V$ value implied by the RCrA color excess data, the first solution seems
the most plausible, in spite of the degeneracy of the fits to the
$A_\lambda^{\rm counts}$ data.

\subsection{Comparing Different Measures of the Dust Optical Depth}
\label{avcomp}

The preceding sections outline a variety of methods used to measure the dust
extinction optical depth of a few representative diffuse interstellar cirrus
clouds: empirical scaling of the 100\micron\ flux after correction for
temperature variations (Sec.~\ref{irdata}), analysis of the background star
count density as a function of apparent magnitude (Sec.~\ref{counts}),
reddening of stellar color distributions (Sec.~\ref{cumcolor}), and shifts of
the stellar color-color distribution (Sec.~\ref{colorcolor}).  The different
methods measure slightly different quantities (as quantified in
Sec.~\ref{clumpy}).  Nevertheless, the methods yield general consistency for
a given cloud.  The full set of $A_V$ estimates, based on both the standard
$R_V=3.1$ value and the preferred $R_V=1.7$ value, are listed in
Table~\ref{av_comp_tbl} and plotted in Figure~\ref{av_comp}.


\begin{deluxetable}{l c c c r c c l}
\small
\tablewidth{473.80984pt}
\tablecaption{Extinction Estimates \label{av_comp_tbl}}
\tablehead{
\colhead{(1)}
  & \colhead{(2)}
  & \colhead{(3)}
  & \colhead{(4)}
  & \colhead{(5)}
  & \colhead{(6)}
  & \colhead{(7)}
  & \colhead{(8)}\\
\colhead{Name} 
  & \colhead{$R_V$}
  & \colhead{Region}
  & \colhead{$A_V^{100\mu\rm m}$}
  & \colhead{$A_V^{\rm counts}$} 
  & \colhead{$A_V^{\rm excess}$}
  & \colhead{$A_V^{UBV}$}
  & \colhead{$A_V^{BVI}$}\\
}
\startdata
RCrA      & 3.1 & on1 & $0.25\pm0.02$ & $0.26\pm0.04$\tablenotemark{a} & $0.10\pm0.02$ & $0.18\pm0.01$ & $0.05\pm0.01$\nl
          & 3.1 & on2 & $0.77\pm0.04$ & $0.61\pm0.04$\tablenotemark{a} & $0.46\pm0.04$ & $0.45\pm0.01$ & $0.36\pm0.01$\nl
          & 1.7 & on1 &               & $0.26\pm0.03$\tablenotemark{a} & $0.07\pm0.01$ & $0.11\pm0.01$ & $0.06\pm0.01$\nl 
          & 1.7 & on2 &               & $0.61\pm0.04$\tablenotemark{a} & $0.27\pm0.02$ & $0.26\pm0.01$ & $0.25\pm0.01$\nl
PV\,1     & 3.1 & on1 & $0.16\pm0.01$ & $0.24\pm0.06$                  & $0.33\pm0.18$ & $0.36\pm0.04$ & $0.28\pm0.04$\nl
          & 3.1 & on2 & $0.35\pm0.01$ & $0.46\pm0.07$                  & $0.65\pm0.18$ & $0.55\pm0.03$ & $0.52\pm0.04$\nl
          & 1.7 & on1 &               & $0.21\pm0.06$                  & $0.20\pm0.11$ & $0.16\pm0.02$ & $0.13\pm0.03$\nl
          & 1.7 & on2 &               & $0.42\pm0.06$                  & $0.37\pm0.10$ & $0.31\pm0.02$ & $0.32\pm0.02$\nl
Paley\,1  & 3.1 & on1 & $0.05\pm0.01$ &                                & $0.16\pm0.06$ && \nl
          & 3.1 & on2 & $0.16\pm0.01$ & $0.12\pm0.23$\tablenotemark{b} & $0.24\pm0.07$ && \nl
          & 1.7 & on1 &               &                                & $0.11\pm0.04$ && \nl
          & 1.7 & on2 &               & $0.09\pm0.17$\tablenotemark{b} & $0.16\pm0.04$ && \nl
Paley\,3  & 3.1 & on1 & $0.06\pm0.01$ &                                & $0.08\pm0.08$ && \nl
          & 3.1 & on2 & $0.14\pm0.01$ & $0.13\pm0.11$\tablenotemark{b} & $0.17\pm0.08$ && \nl
          & 1.7 & on1 &               &                                & $0.06\pm0.05$ && \nl
          & 1.7 & on2 &               & $0.09\pm0.09$\tablenotemark{b} & $0.10\pm0.04$ && \nl
\tablenotetext{a}{$A_V^{\rm counts}$ is determined by fitting the $UBVRI$
  Wolf diagram shifts to a clumpy cirrus cloud model: standard $R_V=3.1$
  extinction law with a fraction $(1-f)=0.1$ of the cloud area covered by
  optically thick dense clumps, or with $R_V=1.7$ and $(1-f)=0.2$
  (Sec.~\ref{clumpy}).}
\tablenotetext{b}{$A_V^{\rm counts}$ is scaled, using the standard $R_V=3.1$
  extinction law or $R_V=1.7$, from $A_U^{\rm counts}$ in the Paley\,1
  ``on2'' region or from $A_U^{\rm counts}$ and $A_B^{\rm counts}$ in the
  Paley\,3 ``on2'' region.  The rest of the ``on2'' and all ``on1''
  $A_\lambda^{\rm counts}$ values are poorly constrained.}
\tablecomments{\\
Col. 4: Derived from 100\micron\ flux (Sec.~\ref{irdata}).\\
Col. 5: Derived from $UBVRI$ star counts/Wolf diagrams (Sec.~\ref{counts}).\\
Col. 6: Derived from $UBVRI$ color excesses (Sec.~\ref{cumcolor}).\\
Col. 7: Derived from $UBV$ color-color diagram (Sec.~\ref{colorcolor}).\\
Col. 8: Derived from $BVI$ color-color diagram (Sec.~\ref{colorcolor}).}
\enddata
\end{deluxetable}

The agreement between different $A_V$ estimates for the PV\,1 cloud is
especially good if one adopts $R_V\lesssim2$ in deriving $A_V$ from the color
excess and color-color data (filled symbols in the second~panel of
Fig.~\ref{av_comp})---the observed trends of $A_\lambda^{\rm counts}$ and
$E(m_\lambda-V)$ versus $\lambda$ for this cloud are, in any case, indicative
of a low $R_V$.  The RCrA ``on2'' region $A_V$ estimates, particularly those
based on $R_V=1.7$ (filled triangles in panel~1 of Fig.~\ref{av_comp}),
display the largest discrepancy: $A_V^{\rm counts}$ is twice as large as
estimates based on the color excess and color-color distribution and is
significantly smaller than the 100\micron\ flux-based estimate,
$A_V^{100\mu\rm m}$.  As discussed in Sec.~\ref{clumpy} above, this is best
explained in the context of the simple two-phase model of a clumpy cirrus
medium.  The dotted line in panel~1 of Figure~\ref{av_comp} indicates a model
designed to match the RCrA ``on2'' observations: $R_V=1.7$, $f=0.8$,
$A_V^{\rm sheet}=0.3$, and $A_V^{\rm counts}=0.6$.  The estimates based on
color excess and color-color data, $A_V^{\rm excess}$, $A_V^{UBV}$, and
$A_V^{BVI}$, correspond to $A_V^{\rm sheet}$.  The counts-based estimate,
$A_V^{\rm counts}$, is boosted by the presence of high density clumps.  The
far infrared flux-based estimate, $A_V^{\rm100\mu{m}}$, is expected to be
larger than both the color- and counts-based estimates and is nominally a
weighted average of $A_V^{\rm sheet}$ and $A_V^{\rm clump}$.  However, since
the characteristic size of the dense clumps is likely to be smaller than the
coarse ($\sim1^\circ$) angular resolution of the far infrared data and since
the dust in the dense clumps is likely to be colder than the dust in the low
density sheet (because of self-shielding), the flux-weighted average
temperature derived by SFD is likely to be biased in favor of the temperature
of the warmer dust in the sheet.  The $A_V$ estimate derived from the scaling
of $F_{100\mu\rm m}^{\rm corr}$ (Sec.~\ref{irdata}) is then lower than the
``true'' area-weighted average optical depth.  The lower and upper~branches
of the dotted line in panel~1 of Figure~\ref{av_comp} show the area-weighted
average $A_V$ for $A_V^{\rm clump}=2.4$ and~3.5, respectively; these bracket
the measured $A_V^{\rm100\mu{m}}$ for RCrA ``on2''.  The actual optical depth
of the dense clumps could be significantly higher, so it is plausible that
stars behind such clumps drop out of the matched $UBVRI$ sample---this is
equivalent to setting $A_V^{\rm clump}\rightarrow\infty$ in the Wolf diagram
analysis (Sec.~\ref{clumpy}).

The ratio of visual optical depth to far infrared brightness,
$A_V/F_{100\mu\rm m}$, is in the range 0.06--0.08 for RCrA (``on1'' and
``on2''), Paley\,1 (``on2''), and Paley\,3 (``on2''), and in the range
0.12--0.13 for PV\,1 (``on1'' and ``on2'').  These results are in good
agreement with other measurements of diffuse interstellar clouds: 0.05--0.1
in \markcite{stark95}Stark's (1995) study of isolated, high-latitude clouds
(including Paley\,1 and Paley\,3); 0.05 in an all-sky survey by
\markcite{boul88}Boulanger \& P{\' e}rault (1988); and 0.04 in a cirrus cloud
in the direction of the Polaris star \markcite{zagury99}(Zagury et~al.\
1999).

The $F_{100\mu\rm m}$ values quoted above for the cirrus clouds in this study
are the calibrated far infrared fluxes from \markcite{schlegel98}SFD, {\it
uncorrected\/} for dust temperature.  Using the temperature-corrected flux,
$F_{100\mu\rm m}^{\rm corr}$ (the flux a cloud would have if it were at the
global average dust temperature of $\langle{T}\rangle=18.0$~K), lowers the
$A_V/F_{100\mu\rm m}^{\rm corr}$ ratio to 0.04--0.06 for RCrA, Paley\,1, and
Paley\,3, and to 0.07--0.09 for PV\,1.  These clouds are colder than the
typical dust cloud in the Galaxy (see Table~\ref{prop_clouds_tbl}), probably
due to the absence of strong heating sources.  Thus, the temperature
correction causes $F_{100\mu\rm m}^{\rm corr}$ to be greater than
$F_{100\mu\rm m}$, and the $A_V/F_{100\mu\rm m}^{\rm corr}$ to be
correspondingly lower.  The typical large dust cloud complex in the
\markcite{boul88}Boulanger \& P{\' e}rault (1988) study is in a hotter
radiation environment than the cirrus clouds in our sample and the
\markcite{zagury99}Zagury et~al.\ (1999) cloud is thought to be heated in
part by the Polaris star; this may explain why their $A_V/F_{100\mu\rm m}$
measurements are at the low end of the range found in this study.  If the SFD
temperature estimates are taken literally, the observed difference between
$A_V/F_{100\mu\rm m}$ for PV\,1 versus the other clouds is not due to
cloud-to-cloud variations in dust temperature, but instead reflects
variations in dust optical properties and/or measurement error.

\subsection{Future Prospects: Sloan Digital Sky Survey Data}
\label{sloan}

The techniques described in this paper are applied to four~representative
cirrus clouds.  However, the application and the results derived from it are
limited by several~factors including: field of view of the observations
($\rm\sim1~deg^2$), photometric depth of the stellar sample ($V_{\rm
lim}\approx{V}_{50}\sim20$), photometric accuracy (typically
$\sigma\gtrsim0.1$~mag), and the cirrus cloud sample size (4).  In the case
of very thin cirrus, a large number of background stars and accurate
photometry are needed to make an accurate determination of the extinction.
This can be accomplished by using a larger field of view and/or increasing
the photometric depth.  Determination of cloud distance can only be achieved
with a substantial number of (bright) foreground stars, and this requires a
wide field of view.  A large sample of optically thin cirrus clouds must be
studied before the conclusions about low $R_V$, small grains, and clumpiness
can be extrapolated to the general diffuse ISM.

The Sloan Digital Sky Survey (SDSS) data set should combine several of the
desired properties outlined above.  The SDSS is an ongoing digital
photometric and spectroscopic survey covering $\rm10^4~deg^2$ in the North
Galactic cap, and a deeper imaging survey in the South Galactic hemisphere
covering $\rm225~deg^2$.  Accurate photometry is being carried out in
five~bands, $u^\prime$, $g^\prime$, $r^\prime$, $i^\prime$, and $z^\prime$
(spanning the entire near-ultraviolet to optical range), complete to limiting
point source magnitudes of $m_{\rm AB}=22.3$, 23.3, 23.1, 22.3, and~20.8,
respectively in the North Galactic cap.  The southern portion of the survey
should extend about 2~mag fainter than the northern portion.  The entire
photometric survey is expected to include over $10^8$ stars and galaxies.

Application of the photometric methods explored in this paper to the SDSS
database should yield a reliable and detailed extinction map, one that is
based on direct measurement as opposed to scaling of far infrared flux, along
with excellent statistics for the measurement of cloud distances.  This
exercise will likely go hand in hand with the development of a new empirical
star count model of the Galaxy, a refinement of the Bahcall-Soneira model.
The SDSS northern sample is expected to be about 2~mag deeper than the sample
analyzed in this study, and this corresponds to about an order of magnitude
increase in the surface density of background objects.  The increased density
of background tracers should translate to an increase in the angular
``resolution'' of the resulting extinction map, especially for the southern
part of the SDSS survey.  The SDSS opens up the prospect of obtaining a
three-dimensional map of cirrus clouds over a quarter of the sky and
conducting an extensive study of diffuse interstellar dust clouds.

\section{Summary}
\label{summary}

We develop techniques for investigating the optical depth, constituent grain
properties, structure, and distance of diffuse interstellar dust clouds.  The
study is based on $UBVRI$ CCD photometry of several thousand stars in
four~$\rm\gtrsim1~deg^2$ fields centered on Galactic cirrus clouds from which
well matched on- and off-cloud samples are constructed.  The main points of
the paper are summarized below:

\begin{itemize}

\item A CCD-based variant of the 75-yr old Wolf diagram method is used to
measure the effective extinction in different bands for the four~cirrus
clouds.  Star counts-based $UBVRI$ extinction curves are presented for
two~clouds, RCrA and PV\,1.  The steepness of the PV\,1 extinction curve
implies $R_V\equiv{A}_V/E(B-V)\lesssim2$, which is unusually low compared to
the standard Galactic value of $R_V=3.1$.

\item A comparison of normalized cumulative stellar color distributions on
and off the cirrus cloud yields color excesses, $E(m_\lambda-V)$ for
$m_\lambda=UBRI$.  The reddening laws for all four~cirrus clouds are steep,
indicative of $R_V$ values which are significantly lower than the standard
value.

\item The optical depth of the RCrA and PV\,1 clouds is also determined from
the shift of the stellar distribution in $U-B$ vs $B-V$ and $B-V$ vs $V-I$
color-color diagrams.  This method depends on an assumed value of $R_V$ which
determines the length, and to a lesser extent, direction of the unit
reddening vector.  Sample incompleteness at the faint end is found to be
unimportant for the color-color analysis.

\item The four~cirrus clouds in this study appear to favor unsually low
values of $R_V$ ($\lesssim2$), comparable to the lowest values found in
previous studies.  This fits in with the clouds' low optical depth ($A_V<1$)
and the general trend of increasing $R_V$ with increasing optical depth (and
vice versa) that is observed for the interstellar medium of the Galaxy.  The
low $R_V$ values in optically thin cirrus clouds is suggestive of a relative
overabundance of small dust grains.

\item Constraints on cloud distances are obtained by studying the cumulative
star counts versus apparent magnitude relations on and off the cirrus in the
context of the Bahcall-Soneira star count model of the Galaxy.  The RCrA
cloud is estimated to lie at a distance of 550$\>$--$\>$750~pc, while the
distance to the PV\,1 cloud is $<1$~kpc.

\item The effects of a non-uniform dust column density are illustrated
through a simple two-phase model: a low extinction sheet with embedded high
density clumps.  A clumpy medium tends to flatten (i.e.,~decrease the degree
of wavelength dependence of) the star counts-based effective extinction
curve.  It also causes the counts-based $A_V$ estimate to be higher than
estimates based on color excess and color-color data and lower than the
estimate based on far infrared flux.  The RCrA Wolf diagram and stellar color
data are readily explained in terms of a non-uniform medium in which 20\% of
the area is covered by dense clumps.

\item Estimates of the dust optical depth derived from $UBVRI$ extinction and
reddening measurements are in good agreement with those based on the
100\micron\ flux.  The values of the $A_V/F_{100\mu\rm m}$ ratios for the
four~clouds are consistent with the ratios found in other studies of diffuse
Galactic dust clouds and follow the expected trends in dust temperature.

\item It should be possible to apply the photometric techniques described in
this paper to the vast Sloan Digital Sky Survey data set in order to carry
out an extensive study of dust properties in the diffuse interstellar medium
and to construct an extinction map.

\end{itemize}

\acknowledgements
A.S.\ and P.G.\ are supported in part by NASA Long Term Space Astrophysics
grant NAG\,5-3232; PG is supported in part by an Alfred P.\ Sloan Foundation
fellowship.  We would like to thank Jean-Philippe Bernard for assistance with
the observations.

\newpage

\clearpage
\newpage

\figcaption[Szomoru.fig01.ps] {Greyscale (negative) representation of the
$V$-band CCD images of the four~cirrus cloud fields.  The bold white contours
indicate the on-cloud (``on1'' and ``on2'') and off-cloud (``off'') regions,
demarcated on the basis of 100\micron\ brightness.  North is up and east is
to the left.  The scale of each image is indicated; the PV\,1 image covers
1.8~deg$^2$, while each of the other three~images covers about 1~deg$^2$.
The optical surface brightness enhancement visible in the on-cloud region
(relative to the off-cloud region) is due to reprocessing (mostly scattering)
of ambient starlight by interstellar dust grains.  The RCrA field is at the
lowest Galactic latitude of the four~fields ($\vert{b}\vert=17^\circ$) and
consequently contains the highest surface density of Galactic field stars.
Areas around bright, badly saturated stars are excluded from data analysis.
\label{v_images}}

\figcaption[Szomoru.fig02.ps] {Cumulative star counts versus calibrated
apparent $U$-band magnitude (Wolf diagrams) in the four~cirrus cloud fields.
The $y$-axis numbers refer to the observed counts in the ``off'' region
(solid black line); the ``on1'' (solid green line) and ``on2'' (solid red
line) counts have been normalized to match the area of the ``off'' region
(the ``on1'' counts are not shown for Paley\,1 and Paley\,3).  The total
number of saturated stars in each field and in each band (for which
photometry is unavailable) is added to the bright end of the cumulative
distribution.  As expected, the ``on2'' (region with highest 100\micron\
brightness) shifts are larger than the ``on1'' shifts.  The dotted blue lines
indicate Bahcall-Soneira star count models for the ``off'' and ``on2''
regions (upper and lower curve, respectively; see Sec.~\ref{distance} and
Fig.~\ref{dist_method}) of RCrA and PV\,1, and for the ``off'' region of
Paley\,1 and Paley\,3.  The overall shape of the observed cumulative star
counts (roughly a power law, with slight curvature) agrees well with that of
the model relations, except for the flattening due to sample incompleteness
at the faint end. The apparent $U$ magnitude at which the sample is 50\%
complete, $U_{50}$, is indicated by a vertical dashed line for each field. 
\label{wolf_u_all}}

\figcaption[Szomoru.fig03.ps] {Same as Figure~\ref{wolf_u_all} for the RCrA
field in the $BVRI$ bands, with Bahcall-Soneira star count models shown only
for the $B$ and $V$ bands.  Note the general trend of decreasing shifts
between on-cloud and off-cloud star count relations as one moves towards
longer wavelengths ($m_\lambda=U{\rightarrow}I$). \label{wolf_bvri_rcra}}

\figcaption[Szomoru.fig04.ps] {The apparent magnitude shift between on- and
off-cloud cumulative star counts, $A_\lambda^{\rm counts}$, plotted as a
function of calibrated apparent magnitude $m_\lambda$ in the $UBVRI$ bands
(light blue, dark blue, green, red, and black lines, respectively) for the
four~cirrus clouds.  The ``on1'' regions of Paley\,1 and Paley\,3 are not
shown as the shifts are only marginally significant.  As expected, the shifts
decrease systematically with increasing wavelength from $U$ to $I$ for a
given cloud/region and tend to be higher in ``on2'' than ``on1''.
\label{wolf_shifts}}

\figcaption[Szomoru.fig05.ps] {Estimate of the uncertainty in the
$A_\lambda^{\rm counts}$ values derived from star counts in the $U$ and $I$
bands (solid and dashed lines, respectively) for RCrA (field with highest
star density) and Paley\,3.  The $\pm1\sigma$ error band, plotted as a
function of calibrated apparent magnitude $m_\lambda$ ($U$ or $I$), is
derived from the statistical uncertainty in the star counts, projected onto
the magnitude axis using the slope $S_\lambda$ of the log[$N$($<m$)]-vs-$m$
relation, combined with the photometric uncertainty.  Poisson error dominates
through most of the plotted apparent magnitude range; photometric error and
incompleteness cause the flaring of the error bands at the faint end.
\label{wolf_shift_err}}

\figcaption[Szomoru.fig06.ps] {Effective extinction curve for RCrA and PV\,1
based on $A_\lambda^{\rm counts}$, the median apparent magnitude (horizontal)
offset between cumulative star counts in ``off'' versus ``on1''/``on2''
regions (squares and triangles, respectively, with $1\sigma$ error bars),
plotted as a function of inverse wavelength.  The $A_\lambda^{\rm counts}$
measurements for the other two~clouds, Paley\,1 and Paley\,3, are not
accurate enough for the computation of an extinction curve.  The solid line
shows the fiducial extinction curve from Cardelli et~al.\ (1989) scaled to
the weighted average value of the visual optical depth $A_V^{\rm counts}$
(see Table~\ref{av_comp_tbl}) for the average Galactic value of
$R_V{\equiv}A_V/E(B-V)=3.1$.  For the RCrA cirrus cloud, the fit is improved
(relative to the uniform dust slab case, $f=1$) if we assume that the
interstellar medium is clumpy with $(1-f)=0.1$ of the cloud area covered by
optically thick dense clumps of dust (dotted line); an identical fit can be
made assuming $R_V=1.7$ and $(1-f)=0.2$.  For the PV\,1 ``on2'' region, the
fit is improved by adopting $R_V=1.7$ (dashed line), which corresponds to the
blue/ultraviolet portion of the dust extinction curve rising more steeply
than for the standard extinction curve ($R_V=3.1$). 
\label{rcra_pv1_ext_curves}}

\figcaption[Szomoru.fig07.ps] {Enlarged section of the $U$-band Wolf diagram
for the RCrA cloud (upper panel) and several realizations of the
Bahcall-Soneira star count model (center and lower panels).  The star counts
in the ``off'', ``on2'', and ``on1'' regions are indicated by bold solid,
dotted, and thin solid lines, respectively, for the observations as well as
the models.  The $U\sim11\>$--$\>$15~mag range shown represents the
transition region between bright apparent magnitudes, for which the bulk of
the stars are in front of the cirrus, and faint apparent magnitudes, for
which most of the stars are in the background.  Star count models are
computed with and without an intervening slab of absorbing material, at
distances of 0.4~kpc and 0.8~kpc that span the range of possibilities (center
and lower panels, respectively), and with extinction corresponding to the
measured $A_U^{\rm counts}=0.89$~mag in ``on2'' (left panel) and $A_U^{\rm
counts}=0.34$~mag in ``on1'' (right panel).  The $1\sigma$ error bars in the
center left panel are derived from an ensemble of model realizations each
using the number of stars appropriate for the actual RCrA ``on2'' area.
\label{dist_method}}

\figcaption[Szomoru.fig08.ps] {Sample cumulative $U-V$ (solid) and $I-V$
(dashed) color distributions of stars in the ``on2'' (thin) and ``off''
(bold) regions of the RCrA and Paley\,3 fields.  The horizontal bar indicates
the median color excess shift measured in each case.  An arbitrary offset (in
magnitudes) has been applied to each $m_\lambda-V$ ($m_\lambda=U$ or $I$)
color scale in each field for convenience in plotting.  \label{cum_color}}

\figcaption[Szomoru.fig09.ps] {Color excess,
$E(m_\lambda-V)\equiv{A}_\lambda-A_V$ ($\lambda=UBRI$), derived from the
median shift in the cumulative stellar color distribution
(Fig.~\ref{cum_color}) of ``on1'' (squares) and ``on2'' (triangles) with
respect to the ``off'' region, plotted as a function of inverse wavelength
for the four~cirrus fields.  A standard $R_V=3.1$ Cardelli et~al.\ (1989)
reddening law scaled to the weighted average value of the visual optical
depth $A_V^{\rm excess}$ (see Table~\ref{av_comp_tbl}) is plotted for the
``on1'' (dashed line) and ``on2'' region (solid line).  The dotted line shows
the best-fit $R_V=1.7$ curve for the ``on2'' region.  Only the ``on2''
errorbars are shown for the sake of clarity.  \label{color_excess}}

\figcaption[Szomoru.fig10.ps] {Color-color diagrams, $U-B$ versus $B-V$ and
$V-I$ versus $B-V$, for stars in the ``off'' (black), ``on1'' (green), and
``on2'' (red) regions of the RCrA and PV\,1 fields (overplotted in this
order).  A shift between the on- and the off-cloud stellar distributions is
visible, most clearly in the case of the RCrA $UBV$ plot.  Theoretical
color-color loci from the Bertelli \etal\ (1994) stellar evolution models are
shown as solid blue lines, averaged over all available ages for
three~different metallicities: $Z=0.0004$, $Z=0.008$, and $Z=0.05$ (top to
bottom).  Reddening vectors corresponding to $A_V=2$~mag are plotted for the
standard $R_V=3.1$ extinction law (red arrow) and for $R_V=1.7$ (black
arrow).  A given offset in the color-color plane translates to a {\it
larger\/} $A_V$ for a standard interstellar reddening law than for an
$R_V=1.7$ law.  \label{color_color}}

\figcaption[Szomoru.fig11.ps] {Illustration of the shift of the on-cloud
[``on1'' (dashed) and ``on2'' (solid)] stellar $UBV$ color-color distribution
with respect to the off-cloud distribution, measured along the $R_V=3.1$
reddening vector, for the RCrA field.  The stellar distributions are compared
in a series of 0.1\,mag-wide slices parallel to the reddening vector, and
$\Delta$ is the offset of each slice in the direction perpendicular to the
reddening vector relative to an arbitrary origin.  The median offset between
``on2''/``on1'' and ``off'' stellar distributions along each slice is
plotted, along with $1\sigma$ error bars based on the width of the
distribution and the number of stars.  The dot-dashed horizontal line
indicates the best-fit value of the visual optical depth, $A_V^{UBV}$,
derived from the weighted mean of the offsets measured in the individual
slices.  \label{shift_ubv}}

\figcaption[Szomoru.fig12.ps] {The effect of clumpiness on the shape of the
counts-based extinction curve, plotted in the normalized form:
$A_\lambda^{\rm counts}/A_V^{\rm counts}$ versus inverse wavelength.  The
calculation is based on a simple two-phase model for the cirrus cloud (see
Sec.~\ref{clumpy}) comprised of a smooth sheet of absorbing material with
optical depth $A_V^{\rm sheet}$ and dense, opaque clumps ($A_V^{\rm
clump}\rightarrow\infty$).  A one-parameter family of model counts-based
extinction curves is obtained by varying the ratio, $A_V^{\rm sheet}/A_V^{\rm
counts}$, for a fixed value of $A_V^{\rm counts}=1$, the overall offset in the
$V$-band Wolf diagram.  The curves shown are computed using the star count
slopes $S_\lambda$ measured in the RCrA field and a standard $R_V=3.1$
extinction law.  The quantity $f$ is area covering factor of the low-optical
depth, ``sheet'' portion of the cloud, and the listed values are based on
$A_V^{\rm counts}=1$.  The higher the degree of non-uniformity of the cloud,
the flatter the shape of the counts-based effective extinction curve; the
limiting shape (for $A_V^{\rm sheet}/A_V^{\rm counts}=0$) is determined only
by the relative values of the $UBVRI$ star count slopes $S_lambda$.
\label{clumpy_ext_curves}}

\figcaption[Szomoru.fig13.ps] {Comparison of $A_V$ estimates derived using
various methods (left to right): scaled from 100\micron\ brightness and far
infrared color temperature measurements (Sec.~\ref{irdata}), cumulative star
counts in the $UBVRI$ bands (Sec.~\ref{counts}), cumulative color
distributions (Sec.~\ref{cumcolor}), and $UBV$ and $BVI$ color-color
diagrams (Sec.~\ref{colorcolor}).  Most error bars do not exceed the symbol
sizes, and, for the sake of clarity, have not been plotted.  There is good
agreement amongst the different $A_V$ estimates.  The dotted line in panel~1
(RCrA ``on2'') shows $A_V$ values for a two-phase clumpy dust model
(Sec.~\ref{clumpy}; Fig.~\ref{clumpy_ext_curves}) in which a fraction
$(1-f)=0.2$ of the area is covered by optically-thick clumps; the range of
$A_V^{100\mu\rm m}$ values (IR) indicated corresponds to
$A_V=2.4\>$--$\>$3.5~mag in these dense clumps.  Open symbols indicate $A_V$
estimates based on the standard $R_V=3.1$ interstellar extinction law; filled
symbols are based on a non-standard extinction law ($R_V=1.7$) which appears
to be favored by the color excess data in all fields.  For Paley\,1 and
Paley\,3, only the ``on2'' estimates are reliable and these are plotted. 
\label{av_comp}}

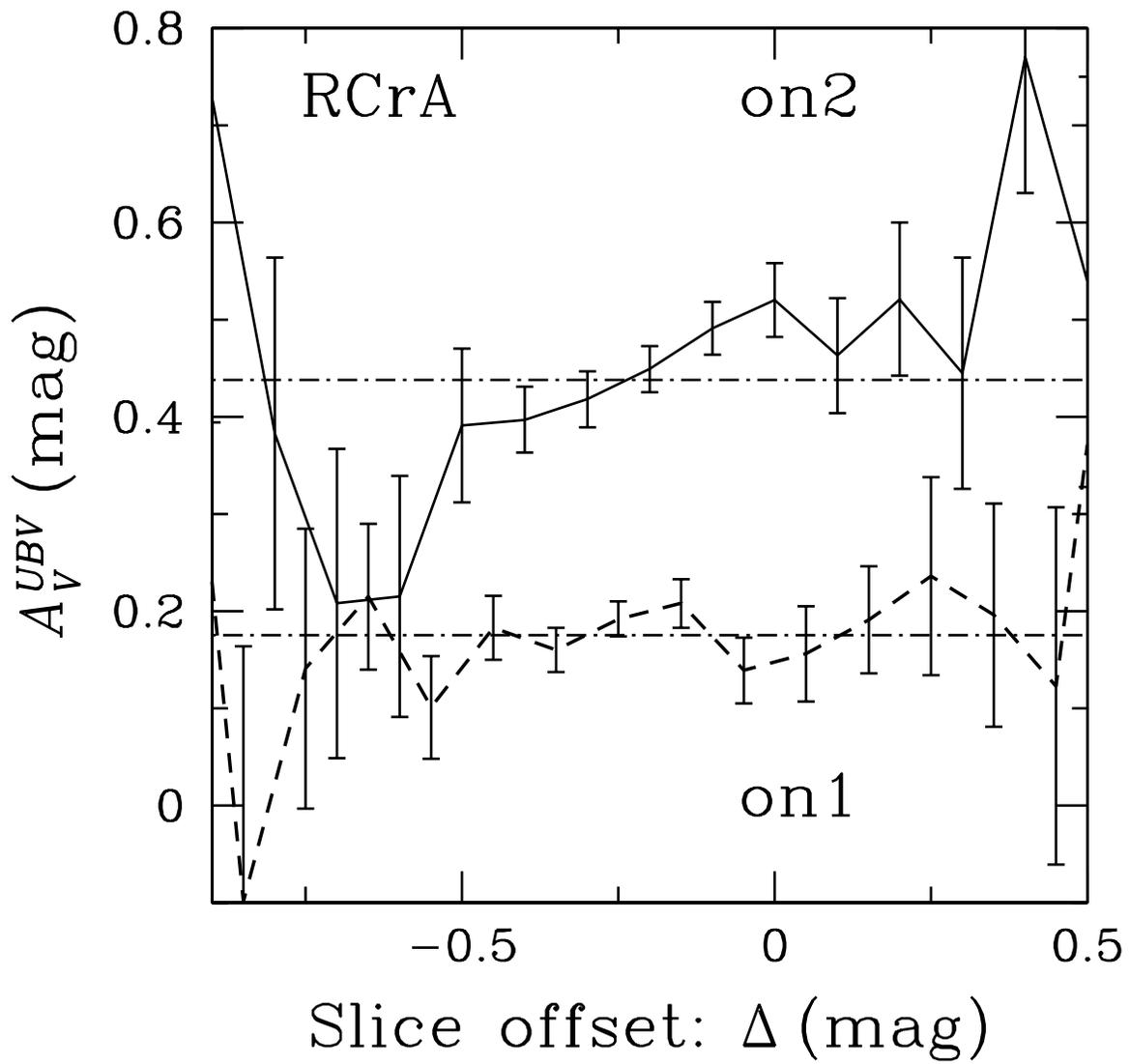
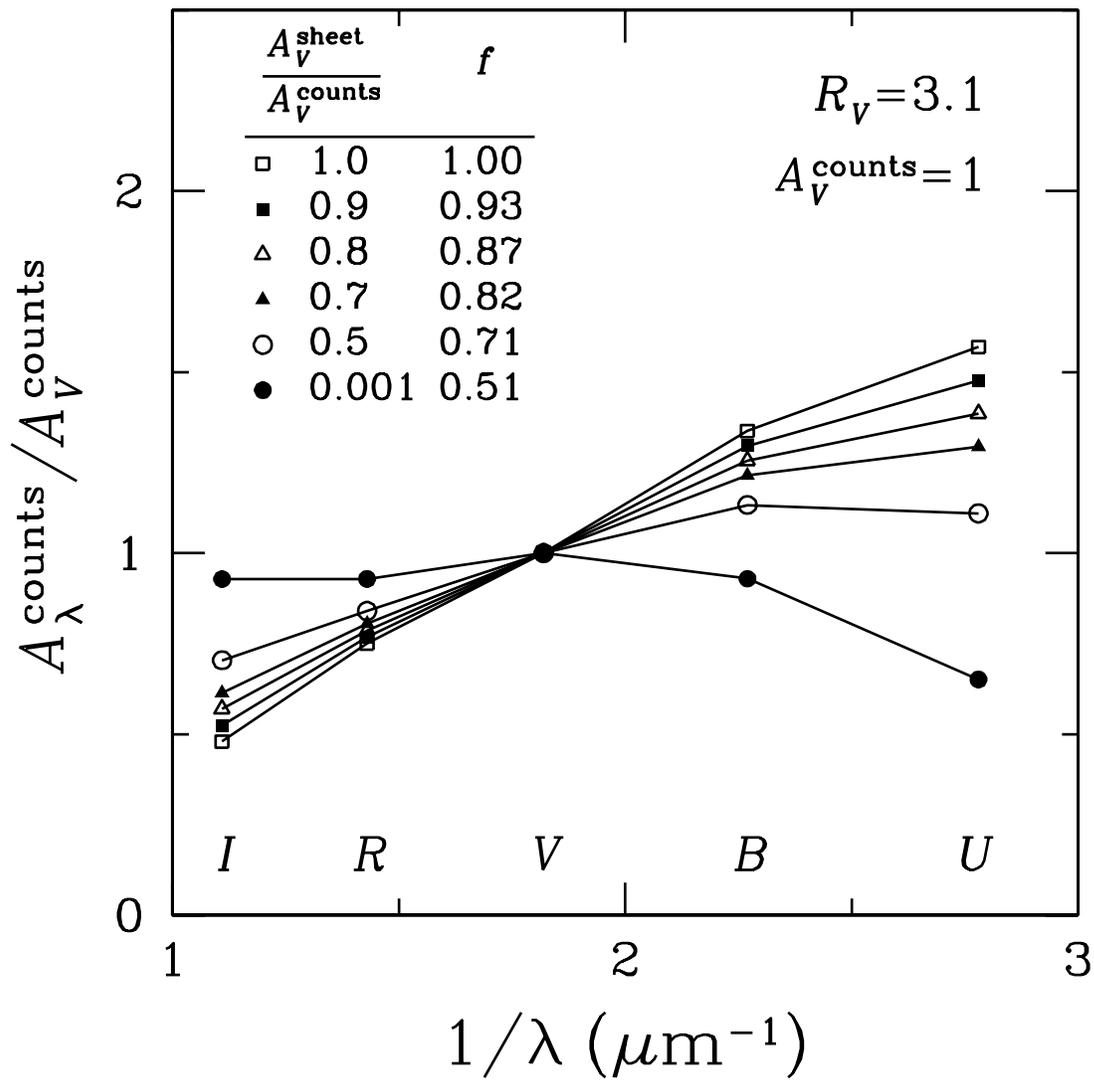
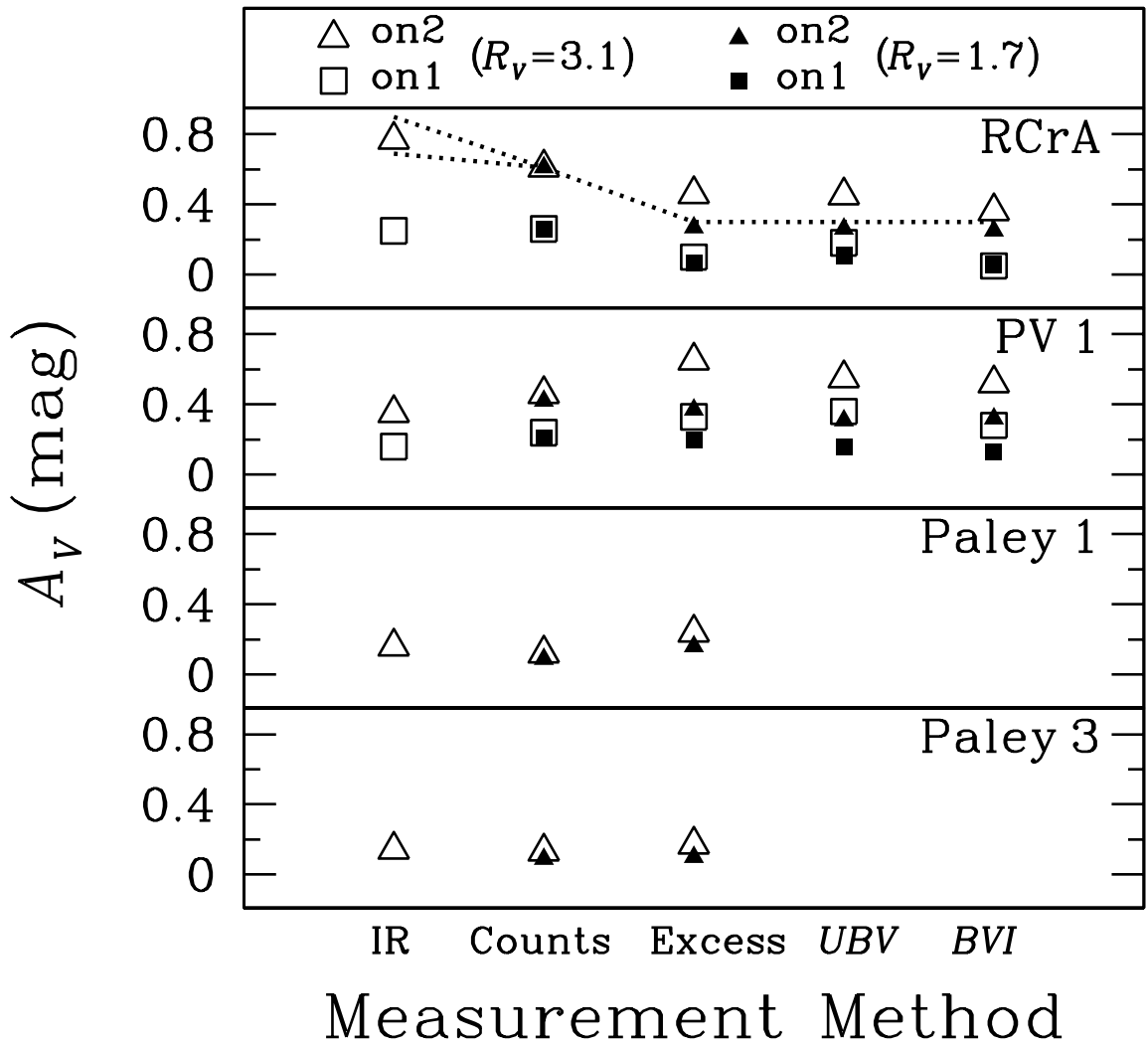
\end{document}